\documentclass[a4paper,11pt]{article}
\usepackage[DIV12]{typearea}
\usepackage{longtable}
\usepackage{booktabs}
\usepackage{amsmath}
\usepackage{amssymb}
\usepackage{bbm}
\usepackage[utf8]{inputenc}
\usepackage{pdflscape}
\usepackage{xspace}
\usepackage[caption=false]{subfig}
\usepackage{nicefrac}
\usepackage{graphicx}
\usepackage{cite}  
\usepackage{nccfoots}
\usepackage{cancel}
\usepackage[small]{caption2}
\usepackage{multirow}

\usepackage{tabulary}
\usepackage{array}
\usepackage{comment}
\usepackage{diagbox}
\usepackage{mathtools}

\newcolumntype{R}[1]{>{\raggedleft\arraybackslash}p{#1}}

\newcommand{\rep}[1]{\ensuremath\boldsymbol{#1}}
\newcommand{\crep}[1]{\ensuremath\bar{\boldsymbol{#1}}}
\newcommand{\x}{\ensuremath\times}
\newcommand{\Z}[1]{\ensuremath{\mathbbm{Z}_{#1}}} 
 
\newcommand{\SO}[1]{\ensuremath{\mathrm{SO}(#1)}}
\newcommand{\SU}[1]{\ensuremath{\mathrm{SU}(#1)}}

\newcommand{\U}[1]{\ensuremath{\mathrm{U}(#1)}}
\newcommand{\E}[1]{\ensuremath{\mathrm{E}_{#1}}}

\newcommand{\bs}[1]{\ensuremath{\boldsymbol{#1}}}

\newcommand{\nphantom}[1]{\sbox0{#1}\hspace{-\the\wd0}}

\usepackage{xcolor}

\definecolor{darkgreen}{HTML}{109930}
\definecolor{pink}{rgb}{0.858, 0.188, 0.478}

\newcommand{\PV}[1]{{\leavevmode\color{red}{#1}}}

\advance \headheight by 3.0truept       
\usepackage[pdftex]{hyperref}

\hypersetup{
    pdftitle = {The landscape of promising non-supersymmetric string models},
    pdfauthor = {Perez-Martinez, Ramos-Sanchez, Vaudrevange}
}
 
\begin{document}

\begin{titlepage}

\begin{flushright}
\normalsize{TUM-HEP 1335/21}
\end{flushright}
\vspace*{1.0cm}

\begin{center}
{\Large\textbf{\boldmath The landscape of promising non-supersymmetric string models\unboldmath}}
\vspace{1cm}

\textbf{%
Ricardo P\'erez-Mart\'inez$^{a}$, Sa\'ul Ramos-S\'anchez$^{b}$, Patrick K.S. Vaudrevange$^c$
}
\\[8mm]
\textit{$^a$\small Facultad de Ciencias F\'isico-Matem\'aticas, Universidad Aut\'onoma de Coahuila,\\ Edificio A, Unidad Camporredondo, 25000, Saltillo, Coahuila, M\'exico}
\\[2mm]
\textit{$^b$\small Instituto de F\'isica, Universidad Nacional Aut\'onoma de M\'exico,\\ POB 20-364, Cd.Mx. 01000, M\'exico}
\\[2mm]
\textit{$^c$\small Physik Department T75, Technische Universit\"at M\"unchen,\\ James-Franck-Stra\ss e 1, 85748 Garching, Germany}
\end{center}

\vspace*{1cm}

\begin{abstract}
Leptoquarks extending the Standard Model (SM) are attracting an increasing attention in the recent 
literature. Hence, the identification of 4D SM-like models and the classification of allowed 
leptoquarks from strings is an important step in the study of string phenomenology. We perform the 
most extensive search for SM-like models from the non-supersymmetric heterotic string 
$\SO{16}\x\SO{16}$, resulting in more than 170,000 inequivalent promising string models from 138 
Abelian toroidal orbifolds. We explore the 4D massless particle spectra of these models in order to 
identify all exotics beside the three generations of quarks and leptons. Hereby, we learn which 
leptoquark can be realized in this string setup. Moreover, we analyze the number of SM Higgs 
doublets which is generically larger than one. Then, we identify SM-like models with a minimal 
particle content. These so-called {\it almost SM} models appear most frequently in the orbifold 
geometries $\Z2\x\Z4\,(2,4)$ and $(1,6)$. Finally, we apply machine learning to our dataset in 
order to predict the orbifold geometry where a given particle spectrum can be found most likely.
\end{abstract}

\end{titlepage}

\newpage

\section{Introduction}

Compactifying six of the ten dimensions of string theory is a crucial step towards four-dimensional 
(4D) string phenomenology. Most of the studies in this context have been performed in theories with 
space-time supersymmetry (SUSY). In particular, toroidal orbifold compactifications of the 
heterotic string~\cite{Dixon:1985jw, Dixon:1986jc,Ibanez:1986tp} have been shown to yield a large 
amount of semi-realistic models~\cite{Ibanez:1987sn,Forste:2006wq,Kobayashi:2004ya,Buchmuller:2005jr,Lebedev:2006kn,Kim:2007mt,Lebedev:2008un,Pena:2012ki,
Nilles:2014owa,Athanasopoulos:2016aws,Olguin-Trejo:2018wpw,Parr:2019bta,Parr:2020oar} 
on which different phenomenological questions can be addressed~\cite{Ko:2005sh,Araki:2007ss,Lebedev:2007hv,Buchmuller:2007zd,Lebedev:2009ag,Ashfaque:2016jha,Carballo-Perez:2016ooy,Olguin-Trejo:2019hxk,Jeong:2019zlr,Kim:2021ngd}. However, since SUSY has not been detected yet, it is worthwhile to entertain the possibility that 
Nature might be described by a non-SUSY model emerging directly from string theory. In this 
sense, the non-SUSY tachyon-free heterotic string with $\SO{16}\x\SO{16}$ gauge 
group in ten dimensions~\cite{Gross:1984dd, Dixon:1986iz, AlvarezGaume:1986jb} can play a special 
role. In order to compare to supersymmetric string models of particle physics, a natural question 
is whether appealing phenomenology can also arise from non-SUSY string theory compactified on SUSY 
preserving spaces.

One can define an orbifold as the quotient of a six-dimensional (6D) torus over a discrete set of 
its isometries, among which the rotational isometries build the so-called point group. There 
are 7103 admissible point groups in six dimensions. 52 of them can leave $\mathcal{N}=1$ 
supersymmetry unbroken in 4D~\cite{Fischer:2012qj}, out of which 17 are Abelian (\Z{N} and \Z{N}\x\Z{M} 
for various orders $M$ and $N$). These 17 point groups give rise to in total 138 inequivalent orbifolds. 
These orbifolds have been the starting point for many studies, trying to connect string theory to 
the supersymmetric extension of the SM. In order to contrast SUSY and non-SUSY phenomenology, in 
this work we focus on orbifold compactifications of the non-SUSY $\SO{16}\x\SO{16}$ heterotic 
string based on those 138 orbifold geometries.

The quest to connect heterotic string theory with non-SUSY particle phenomenology is not new. 
Models with promising properties have been constructed using orbifolds in the bosonic~\cite{Blaszczyk:2014qoa} 
and fermionic formulation~\cite{Ashfaque:2015vta,Faraggi:2020fwg,Faraggi:2020wld,Faraggi:2020hpy}, 
Calabi-Yau manifolds~\cite{Blaszczyk:2015zta}, and so-called coordinate-dependent 
compactifications~\cite{Abel:2015oxa,Abel:2017vos}. Also, string models with spontaneously
broken SUSY have been considered~\cite{Angelantonj:2014dia,Florakis:2016ani,Florakis:2017ecd}.
However, our present study represents the most extensive search up to now for non-SUSY string 
models that reproduce features of the Standard Model (SM) of particle physics. 

By using the \texttt{orbifolder} program~\cite{Nilles:2011aj}, modified to the construction of 
string models without SUSY, we obtain more than 170,000 inequivalent SM-like models from the 138 
orbifold geometries of interest, as summarized in table~\ref{tab:allmodels}. This enormous 
landscape of SM-like models invites to pose questions that may hint towards fruitful corners in the 
landscape where the best phenomenology could emerge. Such questions include:
\begin{itemize}
  \item What kind of exotic matter fields can we obtain from SM-like string models? 
  \item Specifically, which of the leptoquarks introduced in the literature~\cite{Buchmuller:1986zs,Dorsner:2016wpm,Diaz:2017lit} 
        (see also ref.~\cite[sec.~95]{Zyla:2020zbs}) can be realized in this string setting?
  \item Are they useful to tackle some of the open questions of particle physics, such as the 
        question of dark matter or the $g_\mu-2$ puzzle? (See also refs.~\cite{Anchordoqui:2021llp,Anchordoqui:2021vrg} 
        for other string approaches to explain the $g_\mu-2$ discrepancy within D-brane string 
        compactifications.)
  \item Is the origin of the SM encoded in the properties of particular orbifold geometries? 
\end{itemize}
In particular, the questions on leptoquarks are motivated by the recent enhancement of the 
$g_\mu-2$ anomaly, which has triggered a renewed interest in this 
area~\cite{Lee:2021jdr,   
FileviezPerez:2021xfw,    
Angelescu:2021lln,        
Dorsner:2019itg,          
Athron:2021iuf,           
Marzocca:2021azj,         
Nomura:2021oeu,           
Keung:2021rps}.           
Additionally, leptoquarks have been long regarded as viable candidates for dark matter or 
solutions to some other flavor issues~\cite{Cai:2017wry,Choi:2018stw}, though not completely 
free of challenges (see e.g.~\cite{Dorsner:2012nq}).  
We do no attempt to address the phenomenology of stringy leptoquarks. Our work establishes the 
foundation of future endeavors in this direction, which consists in describing what leptoquarks can 
be realized in string constructions. 
In this work, we provide some tools to address these questions by either inspecting systematically 
the properties of the identified models or applying machine learning techniques, as has been done 
recently in the SUSY case~\cite{He:2017aed,Ruehle:2017mzq,Carifio:2017bov,Carifio:2017nyb,Mutter:2018sra,Halverson:2019tkf,Cole:2019enn,Parr:2019bta,Halverson:2020opj,Parr:2020oar,Larfors:2020ugo,Deen:2020dlf,Anderson:2020hux,Ruehle:2020jrk,CaboBizet:2020cse,Bena:2021wyr}.

The content of this work is structured as follows. In section~\ref{sec:classification}, we discuss 
the setting of our search for SM-like string models and provide an overview of our results. In 
section~\ref{sec:exploring}, we analyze the massless spectra of our SM-like models in order to i) 
identify the most promising cases, dubbed here {\it almost SM}, and ii) uncover patterns in string 
theory that may lead to the best phenomenology. General features of the spectra of our models are 
deferred to the appendices. In section~\ref{sec:benchmark}, we illustrate the qualities of our 
models by discussing some properties of a couple of sample models. 
Finally, in section~\ref{sec:conclusions}, we give our conclusions and outlook.

\section{The landscape of non-supersymmetric heterotic orbifolds}
\label{sec:classification}
 
We consider the heterotic string without SUSY in $D=10$ with gauge group $\SO{16}\x\SO{16}$. This 
theory can be obtained from the supersymmetric heterotic string with gauge group $\E8\x\E8$ in the 
bosonic or fermionic formulation~\cite{Dixon:1986iz,AlvarezGaume:1986jb}. The 10D massless spectrum 
of this non-SUSY string theory is tachyon and anomaly free, and consists of 240 gauge bosons, 
256 spinors and 256 cospinors. The dilaton, graviton and Kalb-Ramond field constitute its gravity 
sector. 

In this work, in order to contrast SUSY and non-SUSY compactifications, we focus on the 138 
orbifold geometries classified in ref.~\cite{Fischer:2012qj}. That is, we follow the traditional 
prescription to arrive at 4D models by orbifold compactifications, see e.g.\ 
refs.~\cite{RamosSanchez:2008tn,Vaudrevange:2008sm,Blaszczyk:2014qoa}. In some detail, we define an 
orbifold geometry as the quotient
\begin{equation}
\mathbbm{O} ~=~ \frac{\mathbbm{R}^6}{S}\,,
\end{equation} 
where $S$ is a so-called space group, whose elements are specified as $g=(\vartheta,\lambda)$. The 
so-called twists $\vartheta$ generate a (rotational) point group $P\subset \mathrm{O}(6)$, whereas 
$\lambda$ correspond to translations. Space group elements act hence on the spatial coordinates 
$y\in\mathbbm{R}^6$ of the extra dimensions according to 
\begin{equation}
y ~\stackrel{g}{\mapsto}~ \vartheta\,y+\lambda\,,\qquad g\in S\,.
\end{equation}
In some cases, $\lambda$ is an element of the 6D torus lattice 
$\Lambda = \left\{n_\alpha e_\alpha ~|~ n_\alpha\in\Z{},\ \alpha=1,\ldots,6\right\}$, where 
$\{e_\alpha\}$ is the basis of $\Lambda$. Space group elements with $\lambda\not\in\Lambda$ are 
called roto-translations. In the absence of roto-translations, the orbifold can be defined also as 
$\mathbbm{O}= \mathbbm T^6/P$, where $\mathbbm T^6 = \mathbbm R^6/\Lambda$. It is evident that $P$ 
must then be a symmetry of $\Lambda$. Thus, in general, for each point group there are various 
orbifold geometries, as different $\Lambda$ can have the same point group symmetry. We are 
interested in toroidal orbifolds with and without roto-translations, where $P$ is Abelian. 
The details of the 138 space groups associated with $\Z{N}$ and $\Z{N}\x\Z{M}$ point groups that we 
consider here were systematically obtained in ref.~\cite{Fischer:2012qj} in the context of SUSY 
compactifications. These are equally useful to arrive at consistent 4D models from the
non-SUSY heterotic string $\SO{16}\x\SO{16}$. We shall explore all of them to find 
phenomenologically promising 4D non-SUSY models, which we call SM-like models. 

Each orbifold geometry, characterized by a space group $S$, leads to a myriad of effective field 
theories in 4D, with a given gauge group $\mathcal{G}_\mathrm{4D}$ and massless spectrum of matter 
fields building representations of $\mathcal{G}_\mathrm{4D}$, where we take all fermions to be 
left-chiral. These result from embedding the chosen orbifold geometry into the gauge degrees of 
freedom. These embeddings can be defined by a 16D shift vector $V_i$ for each rotational generator 
of the point group, and up to six 16D Wilson lines $W_\alpha$, $\alpha=1,\ldots,6$, subject to 
consistency conditions including especially modular invariance (see 
refs.~\cite[sec.~3.2]{Blaszczyk:2014qoa} and~\cite{Ploger:2007iq}). Standard techniques yield then 
the gauge groups and massless matter spectra on which we focus in this work.

We define a SM-like model by the following properties of the 4D gauge group 
and massless matter spectrum:
\begin{itemize}
  \item $\mathcal{G}_\mathrm{4D} = \mathcal{G}_{\mathrm{SM}} \x [\mathrm{U(1)}']^n \x \mathcal{G}_\text{hidden}$, 
        where $\mathcal{G}_{\mathrm{SM}} = \SU3_C \x \SU2_L \x \U1_Y$ is the SM gauge group, 
        $\mathcal{G}_\text{hidden}$ is a non-Abelian gauge group, usually built as a product of 
        $\SU{N}$ group factors, and $n>0$ is an integer number subject to the condition that 
        $\mathrm{rank}(\mathcal{G}_\mathrm{4D})=16$. $\mathcal G_\text{hidden}$ is considered 
        ``hidden'' because (almost) none of the SM fields is charged under this group. 
  \item The 4D massless spectrum consists of exactly three generations of chiral fermions for 
        quarks and leptons (including three right-handed neutrinos) and at least one 
        Higgs doublet, a number of exotic fermions that are vector-like with respect to the SM, 
        exotic scalars, and several SM-singlet scalars and fermions. In this way, all 
        exotics can in principle be decoupled without breaking the SM gauge group.
\end{itemize}
The SM hypercharge is non-anomalous and compatible with \SU5 grand unification. In most cases, one 
of the additional $\U1'$s appears anomalous (where the anomaly is canceled by the Green--Schwarz 
mechanism~\cite{Green:1984sg}). Note that an arbitrary number of (vector-like) exotics, Higgs 
doublets and singlets arising from the compactifications are allowed for a SM-like model.

\begin{table}[t!]
\begin{center}
\resizebox{\textwidth}{!}{
\begin{tabular}{|c|c||c|c||c|c||c|c||c|c|}
\hline
orbifold & \# models  & orbifold & \# models & orbifold & \# models & orbifold & \# models & orbifold & \# models\\ 
\hline\hline
$\boldsymbol{\Z2\x\Z2}$ &  & $\boldsymbol{\Z2\x\Z4}$ &  & $\boldsymbol{\Z2\x\Z4}$ & & $\boldsymbol{\Z3\x\Z3}$ & & $\boldsymbol{\Z3}$  & \\ 
(1,1) & 3 & (3,3) & 3935  & (9,3) & 1491 & (4,2) & 17 & (1,1)  & 155  \\ 
(2,1) & 8    & (3,4) & 4779  & (10,1) & 3562 & (5,1) & 595 & $\boldsymbol{\Z4}$  &   \\ 
(3,1) & 10   & (3,5) & 9100  & (10,2) & 3250 & $\boldsymbol{\Z3\x\Z6}$ &  & (1,1)  & 1  \\
(5,1) & 1    & (3,6) & 1916  & $\boldsymbol{\Z2\x\Z6}$-I &  & (1,1) & 43 & (2,1)  & 12  \\
(6,1) & 46   & (4,1) & 1570  & (1,1) & 109 & (1,2) & 2 & (3,1)  & 17  \\
(7,1) & 145  & (4,2) & 6179  & (1,2) & 142 & (2,1) & 98 & $\boldsymbol{\Z6}$-I   &   \\
(9,1) & 6    & (4,3) & 3445  & (2,1) & 45 & (2,2) & 4  & (1,1)  & 33  \\
(10,1) & 5   & (4,4) & 2905  & (2,2) & 310 & $\boldsymbol{\Z4\x\Z4}$  &  & (2,1)  & 31  \\
(12,1) & 10  & (4,5) & 3336  & $\boldsymbol{\Z2\x\Z6}$-II &  & (1,1) & 234 & $\boldsymbol{\Z6}$-II &   \\
$\boldsymbol{\Z2\x\Z4}$ &  & (5,1) & 1771  & (1,1) & 8 & (1,2)  & 90 & (1,1)  &  31 \\
(1,1) & 1779 & (5,2) & 2413  & (2,1) & 156 & (1,3) & 637 & (2,1) & 77  \\
(1,2) & 4590 & (6,1) & 2648  & (3,1) & 150 & (1,4) & 5 & (3,1)  & 140  \\
(1,3) & 3117 & (6,2) &  5542 & (4,1) & 143 & (2,1) & 300 & (4,1)  & 12  \\
(1,4) & 2215 & (6,3) & 3726  & $\boldsymbol{\Z3\x\Z3}$ &  & (2,2) & 172 & $\boldsymbol{\Z8}$-I  &   \\
(1,5) & 9388 & (6,4) & 3574  & (1,1) & 10 & (2,3) & 600 & (1,1)  & 4  \\
(1,6) & 7119 & (6,5) & 1895  & (1,2) & 71 & (2,4) & 2 & (2,1)  & 4  \\
(2,1) & 1587 & (7,1) & 1908  & (1,3) & 3409 & (3,1) & 339 & $\boldsymbol{\Z8}$-II  &   \\
(2,2) & 2886 & (7,2) & 2322  & (1,4) & 1843 & (3,2) & 208 & (1,1)  & 330  \\
(2,3) & 6174 & (7,3) & 612   & (2,1) & 17 & (4,1) & 1665 & (2,1)  & 93  \\
(2,4) & 9283 & (8,1) & 4926  & (2,2) & 521 & (4,3) & 1 & $\boldsymbol{\mathbb{Z}_{12}}$-II  &   \\
(2,5) & 2066 & (8,2) & 3970  & (2,3) & 6402 & (5,1) & 579 & (1,1)  & 102  \\
(2,6) & 4029 & (8,3) & 1919  & (3,1) & 9 & (5,2) & 2 &   &   \\
(3,1) & 2302 & (9,1) & 2346  & (3,2) & 1584 & $\boldsymbol{\Z6\x\Z6}$ &  &   &   \\
(3,2) & 5957 & (9,2) & 464   & (4,1) & 413 & (1,1) & 12 &   &   \\\hline
\end{tabular}
}
\caption{Total number of inequivalent SM-like string models obtained in our extensive search using 
138 orbifold geometries classified in ref.~\cite{Fischer:2012qj}. We find 170,219 inequivalent 
SM-like models in 104 orbifold geometries. In the ``orbifold'' columns we label the considered 
orbifold geometries by their point group (\Z{N} or \Z{N}\x\Z{M}) and the pair $(i,j)$. The latter 
refer to the $i$-th torus lattice and the $j$-th roto-translation element, following the notation 
of ref.~\cite{Fischer:2012qj}. The columns labeled by ``\# models'' display the number of SM-like 
models for the corresponding orbifold geometry.}   
\label{tab:allmodels}
\end{center}
\end{table}

With the goal of performing an extensive search for SM-like models, we use the \texttt{orbifolder}, 
which we adapted to perform these non-SUSY compactifications. The \texttt{orbifolder} creates 
randomly and consistently the essential parameters to construct inequivalent and (perturbatively) 
tachyon-free SM-like models and computes their massless matter spectra.\footnote{Specifically, 
it creates string models by choosing shift vectors and Wilson lines that describe the geometrical 
orbifold action on the string's gauge degrees of freedom, and verifies their consistency under the 
worldsheet modular invariance conditions~\cite{Blaszczyk:2014qoa,Ploger:2007iq}.} Using this tool and exploring 
all 138 orbifold geometries we find SM-like models in 104 out of 138 orbifold geometries. Our 
results are presented in table~\ref{tab:allmodels}. We find a total of 170,219 inequivalent 
promising models, where 169,177 (1,042) belong to the \Z{N}\x\Z{M} (\Z{N}) orbifold geometries. The 
largest number of SM-like models was found in the \Z2\x\Z4 (\Z8-II) orbifold geometries with 147,996 
(423) SM-like models, which reveals a common feature between SUSY and non-SUSY promising orbifold 
compactifications, see e.g.~\cite{Olguin-Trejo:2018wpw,Parr:2019bta,Parr:2020oar}. Our results 
represent, as far as we know, the most extensive search for SM-like models from string theory. Yet 
our search is not exhaustive. In particular, about 1,000 SM-like models with point groups \Z8-I and 
\Z2\x\Z2 were identified before in ref.~\cite{Blaszczyk:2014qoa} and do not appear in our current 
search.

\section{Exploring the SM-like models}
\label{sec:exploring}

\subsection{Vector-like exotics and Higgs doublets}
\label{subsec:vle-H}

We are now interested in knowing explicitly the types of vector-like exotic (VLE) representations 
and the number of Higgses that appear in all 170,219 identified SM-like models. The motivation of 
this study of the particle content is twofold. First, we aim at identifying the most promising 
SM-like candidates, i.e.\ those whose features best fit known observations. Secondly, among the VLE 
matter found in these constructions, inspecting the qualities of the leptoquark sector may be 
relevant for diverse phenomenological questions (see 
e.g.~\cite{Cai:2017wry,Choi:2018stw,Becirevic:2018afm,Dorsner:2019itg,Lee:2021jdr}), including the 
recent enhancement of the muon $g_\mu-2$ anomaly.

Exotic matter refers to representations of ${\cal{G}}_{\mathrm{SM}} = \SU3_C\x\SU2_L\x\U1_Y$ 
appearing in the 4D massless spectrum of an orbifold compactification, beyond the three generations 
of SM fermions, including three right-handed neutrinos, and one Higgs doublet. Further, to be 
characterized as vector-like, i) each exotic fermion must be accompanied by another fermion with 
the exact opposite charges, or ii) it must be a scalar. Abusing of the term, we shall count as VLE 
matter also additional fermionic singlets under $\mathcal{G}_\mathrm{SM}$, and scalar SM-singlets. 
The former may play the role of sterile right-handed neutrinos (see e.g.\ ref.~\cite{Buchmuller:2007zd} 
for its SUSY equivalent), and the latter can be regarded as scalar dark matter candidates in the 
framework of Higgs portals~\cite{Arroyo:2016wal,Arcadi:2019lka}, or also flavon fields in the 
Froggatt--Nielsen mechanism~\cite{Froggatt:1978nt,Babu:2009fd}. We shall refer to the latter simply 
as flavons here.

We report our findings on the different types of exotics in tables~\ref{tab:VLERf}--\ref{tab:AN-es} 
of appendix~\ref{app:tables-exh}. We list all types of VLE matter representations with respect to 
the SM gauge group. We find 26 kinds of VLE fermions and other 26 representations for exotic 
scalars. Tables~\ref{tab:VLERf} and~\ref{tab:VLERs} show the percentage of models that exhibit any
of the different exotic fermion or scalar representations. We observe that fermion and scalar 
singlets are always present in all the models from $\Z{N}$ and $\Z{N}\x\Z{M}$ orbifolds. 
Tables~\ref{tab:AN-VLEf} and~\ref{tab:AN-es} present the average numbers of exotic fermions 
and scalars, respectively.

As a sample case, consider the 155 SM-like models that arise from orbifolds with \Z3 point group. 
The second column of table~\ref{tab:VLERf} shows that the only exotic fermions with standard SM 
quantum numbers are down-type quark singlets and lepton doublets, which appear in most of these 
models. There are on average about three of these states, as we can see in table~\ref{tab:AN-VLEf}. 
Further, about half of the models exhibit many kinds of fractionally charged fermions~\cite{Schellekens:1989qb}. 
Concerning the exotic scalars, in table~\ref{tab:VLERs} we observe that \Z3 SM-like models exhibit 
generically various types of (scalar) leptoquarks. In the notation of ref.~\cite{Dorsner:2016wpm}, 
we identify the leptoquarks $\tilde R_2 : (\rep3,\rep2)_{\nicefrac{1}{6}}$ and 
$\bar S_1 : (\crep{3},\rep1)_{\nicefrac{-2}{3}}$ in about half of the \Z3 SM-like models, and 
$S_1 : (\crep{3},\rep1)_{\nicefrac{1}{3}}$ in all models of this orbifold geometry. We see in 
table~\ref{tab:AN-es} that they are not very abundant in these models: there are on average 
$\sim1.5$ leptoquarks $\tilde R_2$ and $\bar S_1$, while the mean value of the multiplicity of 
$S_1$ leptoquarks is about $5.6$ in these models. 

Interestingly, our tables reveal that the leptoquark scalars $S_1, \bar S_1$ and $\tilde R_2$ 
identified in the \Z3 example are generic in all SM-like orbifold models. No other leptoquarks 
appear. As we shall shortly see, the existence of these leptoquarks might be related to a 
string-specific structure of localized strings in extra dimensions related to an \SU5 grand 
unification, so-called local GUTs.

From tables~\ref{tab:VLERs} and~\ref{tab:AN-es}, we note that there is a large number of scalar 
fields with SM quantum numbers $(\rep1,\rep2)_{\nicefrac{1}{2}}$. These fields correspond to 
Higgs doublets in our SM-like models. Thus, we find different extensions of the SM with various 
numbers (from 1 to 55) of Higgs doublets, such as those previously studied from a bottom-up 
perspective~\cite{Barroso:2006pa,Ferreira:2008zy,Ivanov:2011ae}. For the different point groups of 
our orbifold geometries, we display in table~\ref{tab:numbH} the number of Higgses we find in all 
170,219 SM-like models. We see that only the \Z6-I orbifold geometries yield models with just one 
Higgs doublet (in 13 out of 64 models). There are 3,192 SM-like models with two Higgs doublets 
distributed in the \Z8-II, \Z2\x\Z4, \Z2\x\Z6-I and \Z3\x\Z3 orbifold geometries. Higher 
multiplicities of Higgs doublets seem favored in our constructions, where most models are endowed 
with 11, 9 or 15 Higgs doublets (20,377, 16,657 and 16,484 SM-like models in each case). Although 
no extra Higgs fields have been observed, they might have interesting implications especially for 
dark matter and Higgs phenomenology~\cite{Chang:2017gla, Keus:2013hya, Alakhras:2017bbe}, and 
for an explanation of the $g_\mu-2$ tension~\cite{Han:2021gfu,Chen:2021jok,Ferreira:2021gke,CarcamoHernandez:2021qhf,Jueid:2021avn}.

As we show in appendix~\ref{app:correlations}, there are high correlations between the numbers of 
different VLE representations appearing in our matter spectra. Especially, we find almost perfect 
correlations among the scalar leptoquarks $\bar S_1$, $\tilde R_2$ and the charged scalars 
$(\rep1,\rep1)_1$. Further, the appearance of leptoquarks $S_1$ and extra Higgs doublets 
$\phi:(\rep1,\rep2)_{\nicefrac{1}{2}}$ is correlated, too. We note that a scalar $\rep{10}$-plet of 
\SU5 precisely decomposes as $\rep{10}=\tilde R_2\oplus\bar S_1\oplus(\rep1,\rep1)_1$, while a 
scalar $\rep5$-plet is built by $\rep5=S_1^*\oplus \phi$, where $S_1^*$ is the complex conjugated 
scalar\footnote{Here, we use that scalars have no chirality. Thus, it is just a matter of 
convention to denote the scalar field that originates from the string in the representation 
$(\crep{3},\rep1)_{\nicefrac{1}{3}}$ either as $S_1$ or as $S_1^*$, such that either $S_1^*$ or 
$S_1$ transforms as the complex conjugated representation $(\rep{3},\rep1)_{\nicefrac{-1}{3}}$.}. 
Thus, the observed correlations suggest a common origin for these fields. It is known that 
heterotic orbifolds \PV{can} produce so-called local GUTs~\cite{Forste:2004ie,Buchmuller:2004hv,Ratz:2007my,Nilles:2014owa}. 
In these scenarios, the gauge symmetry is enhanced to an \SU5 GUT not in 4D, but locally in extra 
dimensions at the orbifold singularities, where full GUT multiplets are realized. Hence, the number 
of leptoquarks seems to be related to these local GUTs.

Finally, we compare our findings on VLE with previous results in the context of MSSM-like models 
that result from heterotic orbifold compactifications~\cite{Parr:2020oar}. We see a few 
differences. The most evident difference is that, in general, SUSY models seem to produce more 
exotic representations than our SM-like models. 
However, exotic fermions with charges $(\crep3,\rep2)_{\nicefrac{5}{6}}$ show up in our models 
(cf.\ table~\ref{tab:VLERf} of appendix~\ref{app:tables-exh}), but do not appear in MSSM-like 
orbifold models. In contrast, we observe several similarities between the massless spectra of 
MSSM-like and SM-like string models: there are roughly between 50 and 200 of SM singlets 
(neutrinos and flavons); the most common exotics with SM quantum numbers are 
$(\crep3,\rep1)_{\nicefrac13}$ and $(\rep1,\rep2)_{\nicefrac{-1}{2}}$, suggesting a local GUT 
picture with $\crep5$-plets of \SU5 localized at some orbifold singularities, as mentioned before; 
following the classification of ref.~\cite{Dorsner:2016wpm} and considering SUSY breakdown in 
MSSM-like models, the only possible leptoquarks in all semi-realistic models are just $S_1$, 
$\bar S_1$ and $\tilde R_2$; and the most common fractionally charged fields have SM charges 
$(\rep1,\rep1)_{\nicefrac12}$ and $(\rep1,\rep2)_{0}$.

\subsection{\emph{Almost SM} models from heterotic orbifolds}
\label{sec:almostSM}

We would like now to identify in the landscape of non-SUSY heterotic orbifolds the models that 
best reproduce the particle content of the SM. With this purpose, we inspect systematically the 
spectra of our models to select those that contain three SM generations of fermions and the 
standard Higgs doublet, along with the least amount of exotic matter. Since SM singlets could 
play an important phenomenological role either as extra sterile neutrinos if they are fermions, 
or flavons or dark matter if the are scalars, we shall not count them as exotic states here.

The special SM-like models whose massless matter spectra display the closest resemblance with 
the SM are called here {\it almost SM}. These models can be classified in two categories:
\begin{itemize}
\item \emph{Models with no exotic fermions}. We find 45 {\it almost SM} of this kind, distributed 
      in the orbifold geometries \Z2\x\Z4 (1,6), \Z2\x\Z4 (2,4) and \Z3\x\Z3 (1,4), as summarized 
      in table~\ref{tab:AN-ASM}. The spectra of these models include, besides the three SM 
      generations, only $S_1$ leptoquark scalars accompanied by various numbers of extra Higgs 
      doublets, right-handed neutrinos and SM-singlet scalars. As we see in table~\ref{tab:numbH-ASM}, 
      we find that there are only three of these models with the minimal number of Higgs doublets 
      (six) in this category.

\item \emph{Models with no exotic scalars}. There are 502 {\it almost SM} models of this category, 
      distributed in ten different \Z2\x\Z4 orbifold geometries. As displayed in table~\ref{tab:AN-ASM}, 
      the scalar sector of their spectra include six Higgs doublets (one of them would be the 
      standard Higgs), and SM-singlet scalars. In the fermionic sector, beyond several right-handed 
      neutrinos, the only exotic fermions have quantum numbers $(\crep3,\rep1)_{\nicefrac{1}{3}}$ 
      and $(\rep1,\rep2)_{\nicefrac{-1}{2}}$ (plus their complex conjugates), mostly originated 
      from full multiplets of \SU5 local GUTs.
\end{itemize}
Additional details, such as the shift vectors and Wilson lines of these selected 547 {\it almost
SM} models, can be found in our website~\cite{website:2021} in a format compatible with the 
(non-SUSY) \texttt{orbifolder}.

Some comments are in order. First, there is no SM-like model without exotics. Second, the existence 
of a few exotics in the models identified as {\it almost SM} might be phenomenologically 
challenging. For example, the $S_1$ scalar leptoquarks of models without exotic fermions could lead 
to rapid proton decay if their couplings with first-generation quarks are unsuppressed and they do 
not develop very large masses. Fortunately, in principle, all leptoquarks in this case and all 
exotic fermions in the second category of {\it almost SM} models can be decoupled,
when some flavons attain vacuum expectation values (VEV). The details of this mechanism are beyond 
the scope of this paper and shall be discussed elsewhere.

As a last comment, let us mention the possibility of SM-like models with only one Higgs doublet, 
see ref.~\cite{Blaszczyk:2014qoa}. We only find seven (six) models of this type, arising from 
the \Z6-I (1,1) (\Z6-I (2,1)) orbifold geometries. Unfortunately, as shown in table~\ref{tab:AN-ASM}, 
these models include several VLE. In the scalar sector, beside 71 flavon or dark matter candidates 
on average, they include two $S_1$ leptoquarks and the exotic representations 
$(\rep3,\rep1)_{\nicefrac{1}{6}}$, $(\rep1,\rep2)_0$ and $(\rep1,\rep1)_{\nicefrac{1}{2}}$. In the 
fermionic sector, in addition to about 200 right-handed neutrinos, we see extra vector-like pairs 
of down-type quark singlets and lepton doublets, as well as fractionally charged exotics in the 
representations $(\rep3,\rep1)_{\nicefrac{1}{6}}$, $(\rep1,\rep2)_0$ and 
$(\rep1,\rep1)_{\nicefrac{1}{2}}$ (paired up with their complex conjugates). Although the exotics 
could in principle be decoupled from the low-energy effective theory, these models seem to be in 
worse shape than our {\it almost SM}.

\subsection{Predicting the stringy origin of the SM with machine learning}
\label{sec:ML}

The previous observations based on a systematic, though limited, search reveal some of the general 
properties of the matter spectra of a subset of all possible SM-like string models. 
Fortunately, by using machine learning (ML) techniques, as in ref.~\cite{Parr:2020oar}, we can 
learn more.

Here, based on the information provided by the identified models, we obtain an ML algorithm that 
predicts the specific orbifold geometry that most likely hosts a SM-like string model with a given 
particle content of exotics. We address this task using supervised machine learning and evaluate 
the quality of our algorithms using the accuracy and the f1-macro. The accuracy of our predictive 
ML algorithm is given by the number of correct predictions divided by the total number of 
predictions. On the other hand, the f1-macro is computed as the average of the f1-scores of each of 
the 104 orbifold geometries. Since our dataset is imbalanced, the f1-macro is more suitable for our 
task, see for example section 3.1 in ref.~\cite{Parr:2020oar}.

In order to compare to the accuracy and the f1-macro of a ``good'' ML algorithm, we first compute 
the so-called null value. The null value is based on the trivial algorithm that always predicts the 
orbifold geometry that appears most frequently in our dataset of 170,219 inequivalent SM-like 
models, independently of the given particle spectrum. In our dataset, \Z2\x\Z4 (1,5) is the 
orbifold geometry that yields the largest number of SM-like models, with a total number of 9,388 
SM-like models. Hence, we can estimate the accuracy of the trivial algorithm based on our dataset: 
it gives a correct prediction with a probability of $5.5\%$, i.e.\
\begin{equation}
\mathrm{null\ value\ accuracy\ :}\qquad\frac{9,388}{170,219} ~\approx~ 5.5 \%\;.
\end{equation}
In addition, we compute the f1-scores and the f1-macro for the trivial algorithm. We obtain 
$\mathrm{f}1(\mathbbm{O})=0$ for $\mathbbm{O} \neq \Z2\x\Z4$ (1,5) and $\mathrm{f}1(\mathbbm{O})=1$ 
for $\mathbbm{O} = \Z2\x\Z4$ (1,5). Hence,
\begin{equation}
\mathrm{null\ value\ f1}\text{-}\mathrm{macro\ :}\qquad\frac{1+0+\ldots+0}{104} ~\approx~ 1 \%\;.
\end{equation}
These are our null values against which we will compare our results in the following.

Before we discuss our ML algorithm, we first split our dataset into $80\%$ training data and $20\%$ 
test data. Since the label of the orbifold geometry is a categorical data (i.e.\ data without 
ordering), we use a one-hot encoding for the labels of the 104 orbifold geometries that host 
SM-like string models. Furthermore, for each SM-like string model, we represent the exotic particle 
spectrum by a 52-dimensional vector of integers (for the 26 fermionic and 26 bosonic exotics as 
listed in tables~\ref{tab:VLERf} and~\ref{tab:VLERs}). Hence, our ML algorithm $f_\mathrm{ML}$ 
takes a 52-dimensional vector $X\in\mathbbm{N}^{52}$ as input (corresponding to the particle 
spectrum of exotics) and gives a 104-dimensional vector as output (corresponding to the one-hot encoded 
prediction of the orbifold geometry that most likely can produce the given particle spectrum),
\begin{equation}
\mathrm{particle\ spectrum\ of\ exotics}\ X ~\xmapsto{~f_\mathrm{ML}~}~ \mathrm{orbifold\ geometry}\ \mathbbm{O}\;. 
\end{equation}

\begin{figure}[t!]
\begin{center}
\subfloat[]{\label{fig:nnaccuracies}
\includegraphics[width=0.5\linewidth]{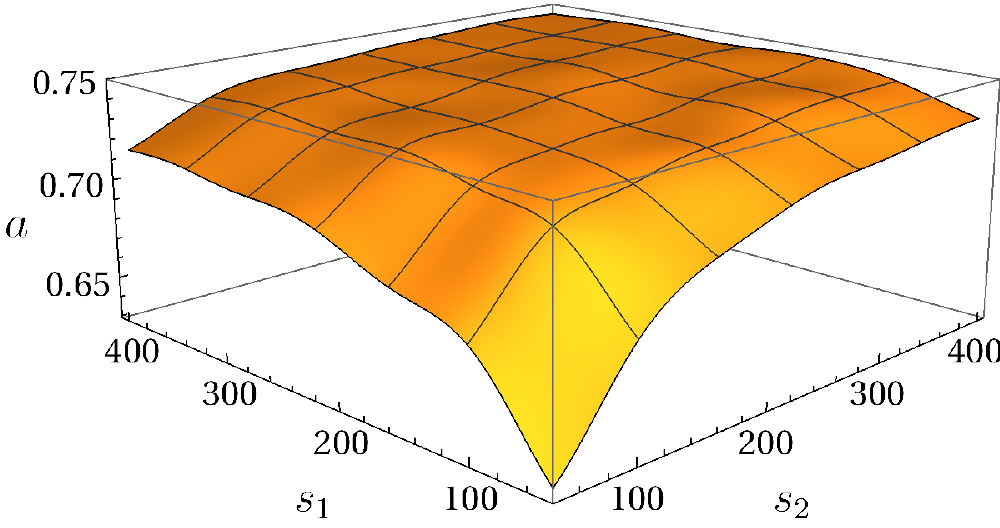}}
\subfloat[]{\label{fig:nnparameters}
\includegraphics[width=0.5\linewidth]{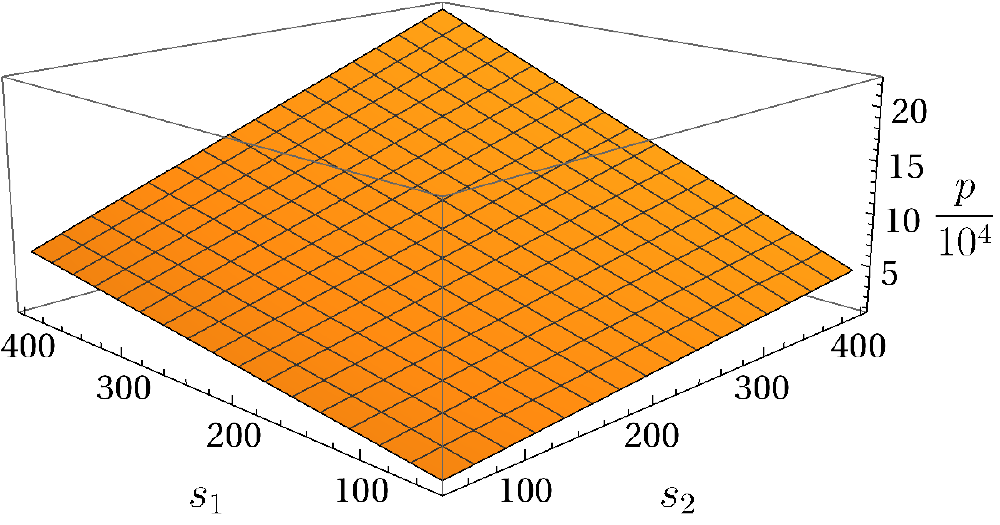}}
\end{center}
\vspace{-0.6cm}
\caption{(a) Accuracies $a$ of the validation set and (b) number of trainable parameters $p$ for 
neural networks with two hidden layers of sizes $s_1 \in\{50,100,\ldots,400\}$ and 
$s_2\in\{50,100,\ldots,400\}$.}
\end{figure}

\begin{figure}[t!]
\begin{center}
\subfloat[]{\label{fig:nnloss}
\includegraphics[width=0.5\linewidth]{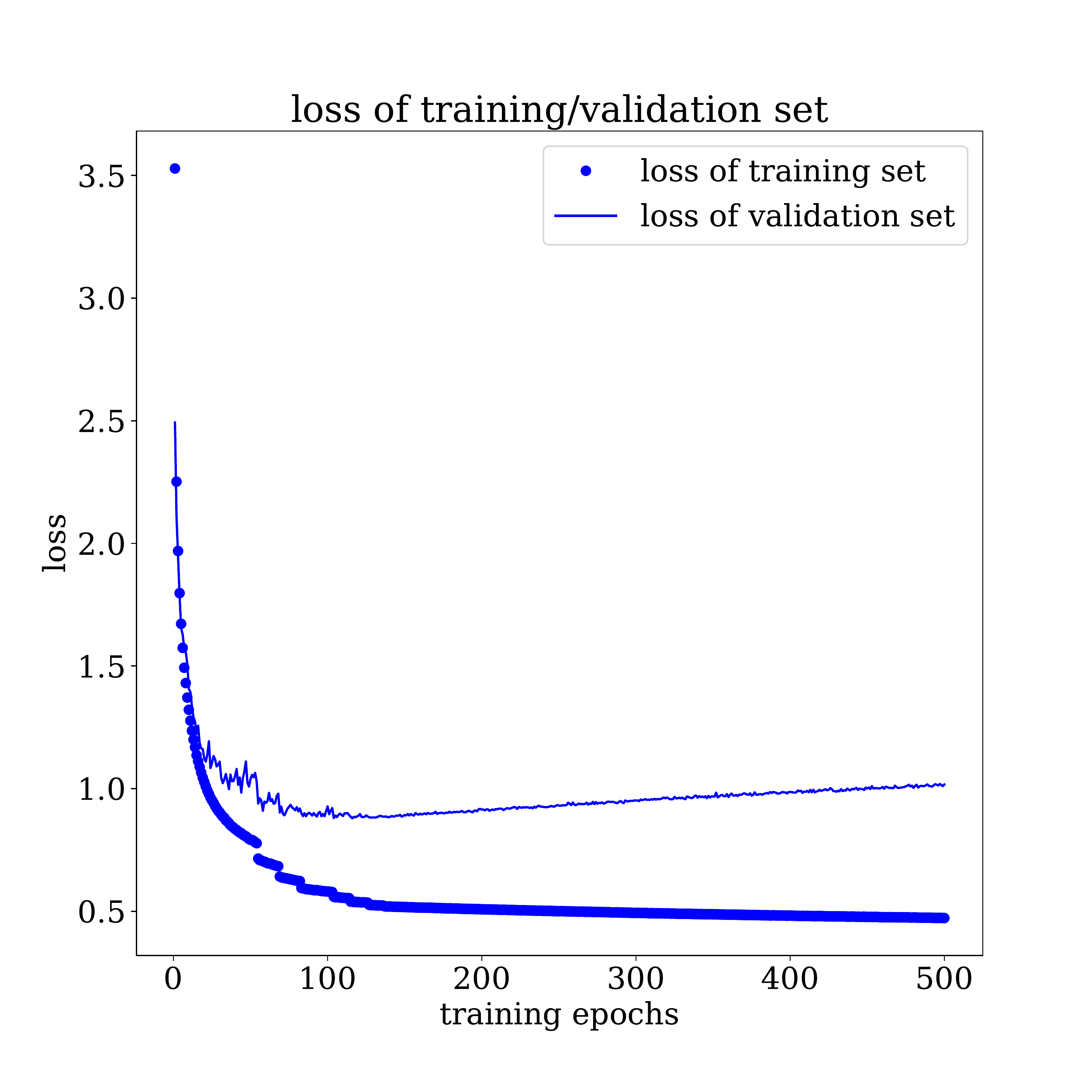}}
\subfloat[]{\label{fig:nnconfusion}
\includegraphics[width=0.5\linewidth]{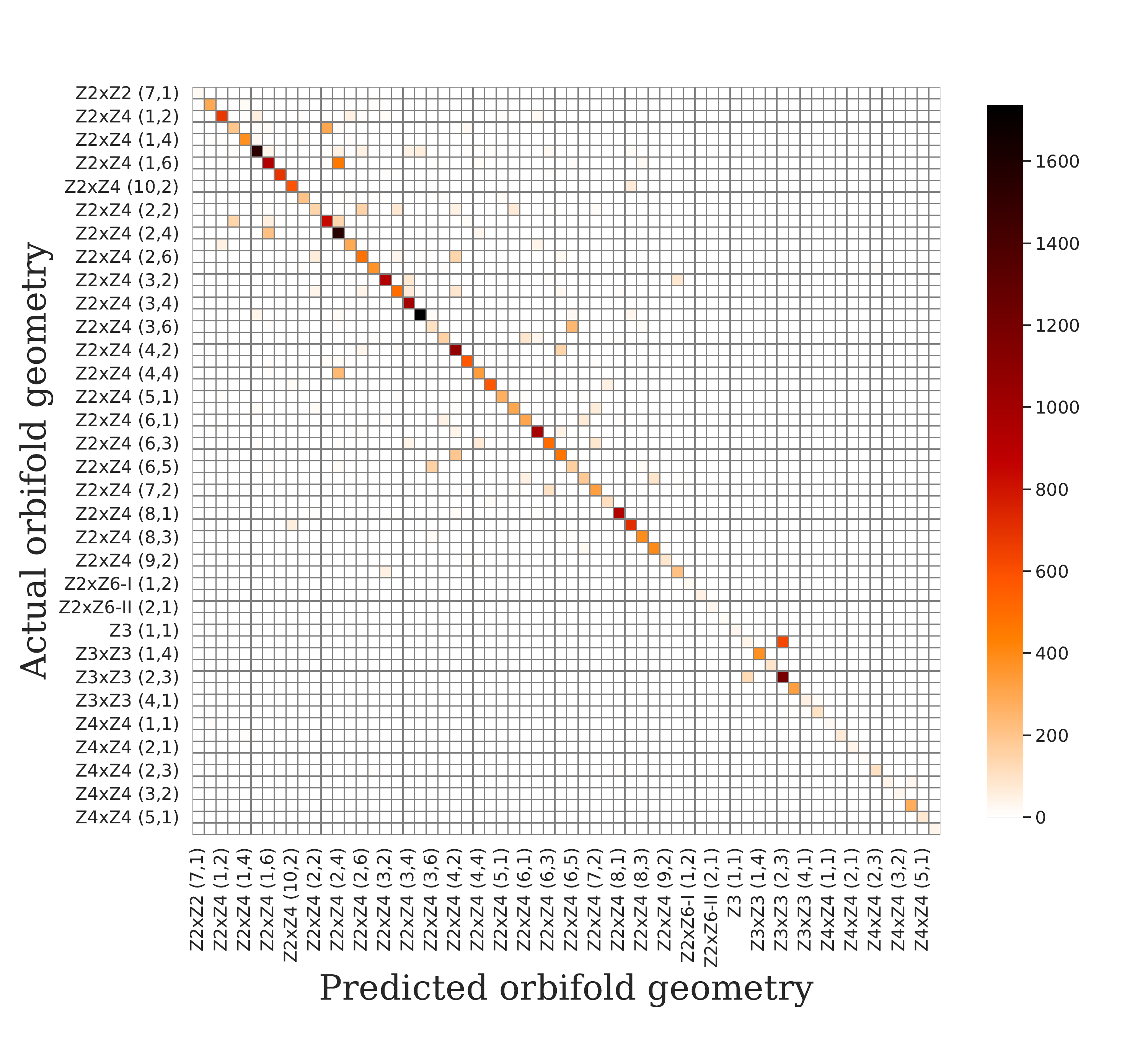}}
\end{center}
\vspace{-0.6cm}
\caption{(a) Loss of training and validation sets after certain number of training epochs of our best 
neural network architecture with $s_1 = 100$ and $s_2 = 350$. Note that the loss of the validation set 
increases after $\approx 120$ epochs, which signals overfitting. (b) Using a majority vote of 21 
neural networks of the size $s_1 = 100$ and $s_2 = 350$, we compute the confusion matrix of the 
test set and display it as a heat map (restricted to those orbifold geometries that have at least 
50 SM-like models in the test set).}
\end{figure}

As ML algorithm, we take a fully connected neural network. The input layer has 52 nodes 
corresponding to $X$, and the output layer has 104 nodes (corresponding to the 104 orbifold 
geometries). We add two hidden layers with $s_1$ and $s_2$ nodes, respectively. Then, the number of 
trainable parameters of the neural network is given by
\begin{equation}
\# ~=~ 53\, s_1 + 105\, s_2 + s_1\,s_2 + 104\;,
\end{equation}
see figure~\ref{fig:nnparameters}, and we want to balance between the accuracy of our neural network 
and the number of trainable parameters. As activation functions we choose ``selu'' except for the 
output layer. There, we use the ``softmax'' activation functions, such that each value in the output 
layer lies in the range $[0,1]$ and the sum of output values is normalized to $1$. Then, we can 
interpret the $i$-th output value as the probability that the $i$-th orbifold geometry can 
reproduce the given particle spectrum. In addition, we use a learning rate of $0.001$ and the loss 
is computed using ``categorical\_crossentropy''. Using our training set, we scan over network 
architectures with
\begin{equation}
s_1 ~\in~ \{50,100,\ldots,400\} \quad\mathrm{and}\quad s_2 ~\in~ \{50,100,\ldots,400\}\;,
\end{equation}
using a $20\%$ validation split and train three times each neural network for 200 training epochs. The 
averaged maximal accuracies of the validation set are evaluated and plotted in 
figure~\ref{fig:nnaccuracies}. The best accuracy of the validation set is around $75\%$ for a network 
architecture with $s_1 = 300$ and $s_2 = 400$ (with $ 178,004$ trainable parameters). However, 
using $s_1 = 100$ and $s_2 = 350$ (with $77,154$ trainable parameters) we already 
obtain an accuracy of $74\%$ (and the f1-macro of the validation set is $73\%$). Thus, we choose 
the smaller but almost equally good network architecture. After $\approx 120$ epochs of training 
the loss of the validation set starts to increase, see figure~\ref{fig:nnloss}. Hence, the neural 
network begins to overfit. Thus, we stop training after 120 training epochs. Then, we construct and 
train 21 neural networks of this architecture and use a majority vote of the 21 individual 
predictions to obtain a final prediction. By doing so, the accuracy of the test set (consisting of 
$20\%$ of all data) increases slightly to $76\%$. We display the confusion matrix of the test set 
as a heat map in figure~\ref{fig:nnconfusion}.

\begin{figure}[t!]
\begin{center}
\includegraphics[width=0.7\linewidth]{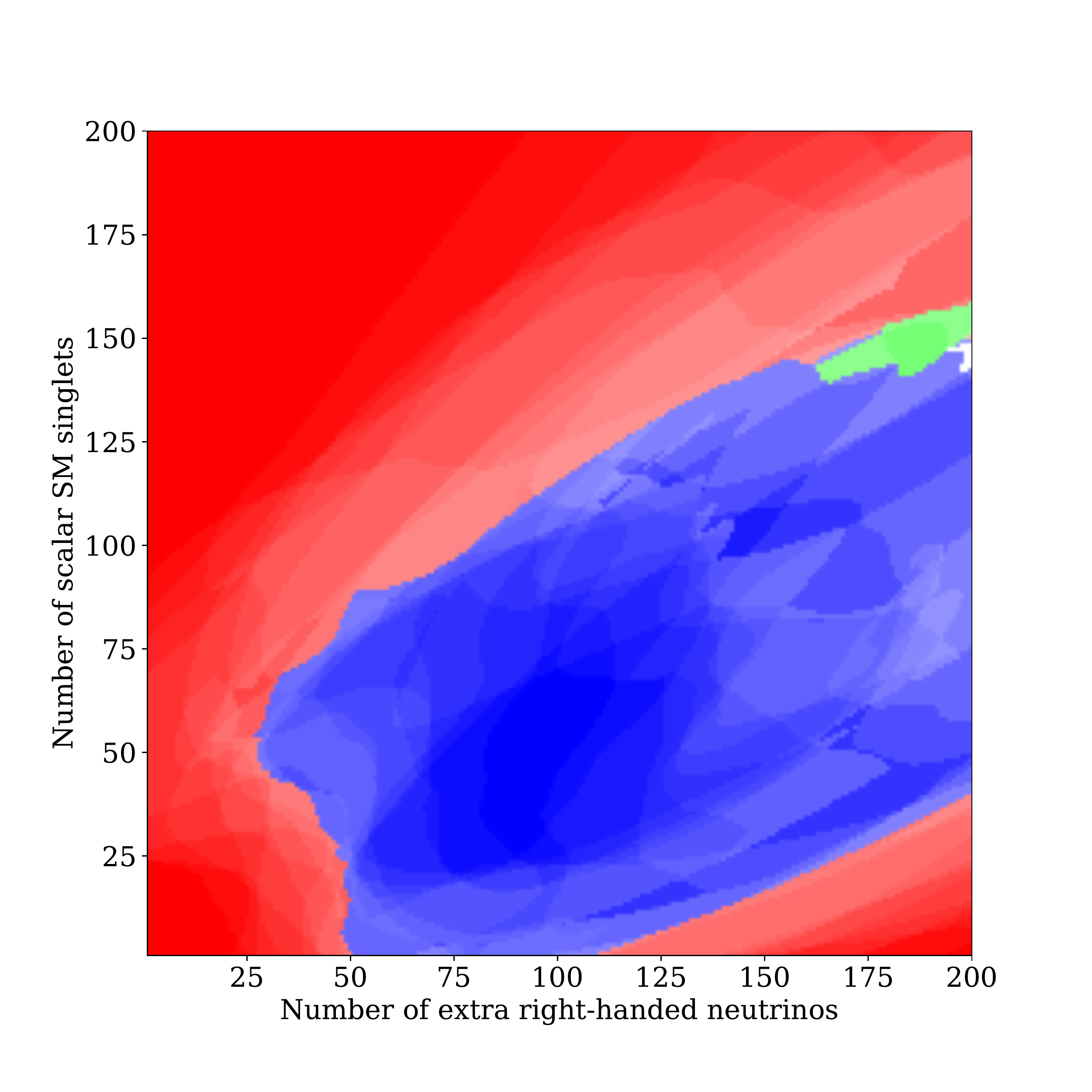}\\%
\vspace{-0.7cm}
\phantom{ttttttt}\includegraphics[width=0.7\linewidth]{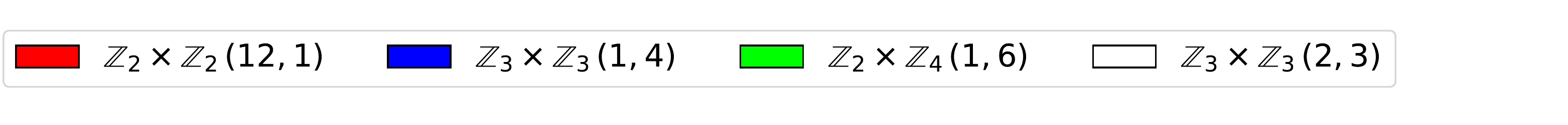}
\end{center}
\vspace{-0.8cm}
\caption{Predictions of orbifold geometries for SM-like spectra without exotics, except for 
$[1,\ldots,200]$ additional right-handed neutrinos and $[1,\ldots,200]$ scalar flavons 
$(\rep1, \rep1)_0$. The transparency of each pixel indicates the accuracy of the respective 
prediction: the less transparent the color, the more accurate the prediction.}
\label{fig:predictions2}
\end{figure}

Now, we can use our trained neural networks to extrapolate to SM-like models that have not been 
discovered in the string landscape so far. By giving a spectrum of exotics to the trained neural 
networks, we obtain a prediction for the orbifold geometry that most likely can host this model. 
For example, we ask the networks what the orbifold geometry is that can most likely reproduce the 
exact SM spectrum without charged exotics. In detail, we specify a SM spectrum that contains in 
addition to the Higgs and the three generations of quarks and leptons only SM singlets: a (large) 
number of right-handed neutrinos (which can be utilized for an extended seesaw mechanism, see 
ref.~\cite{Buchmuller:2007zd}) and a (large) number of SM scalar singlets. The results are 
visualized in figure~\ref{fig:predictions2}. For certain numbers of singlets the orbifold 
geometries \Z2\x\Z2 (12,1), \Z3\x\Z3 (1,4) or \Z2\x\Z4 (1,6) are 
predicted to be able to reproduce these spectra. 

\begin{figure}[h!]
\begin{center}
\begin{tabular}{cc}
(a) & \begin{minipage}{\textwidth}\includegraphics[width=\linewidth]{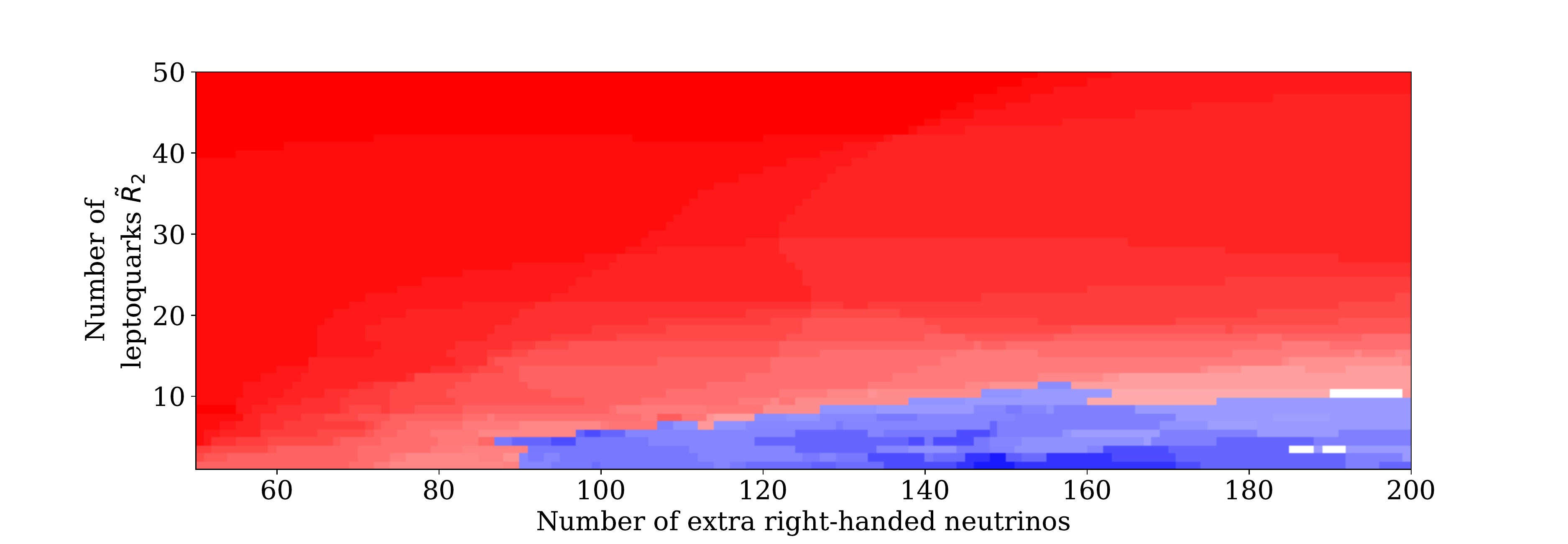}\end{minipage}\\[-2mm]
(b) & \begin{minipage}{\textwidth}\includegraphics[width=\linewidth]{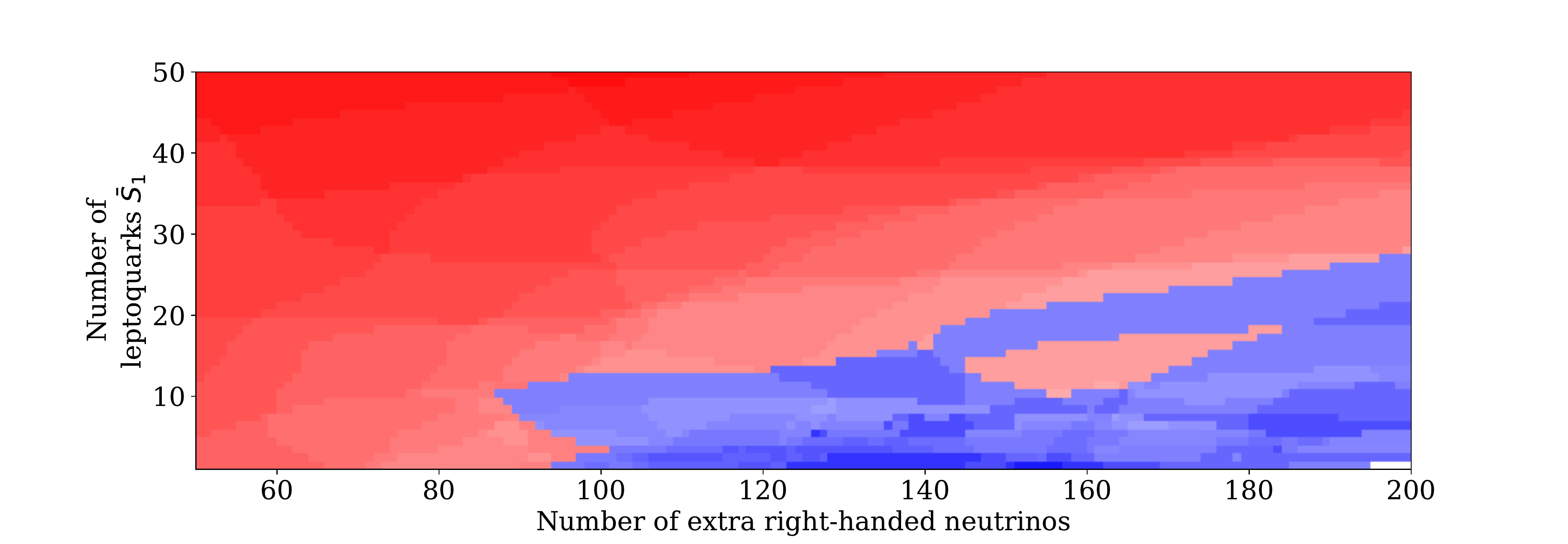}\end{minipage}\\[-2mm]
(c) & \begin{minipage}{\textwidth}\includegraphics[width=\linewidth]{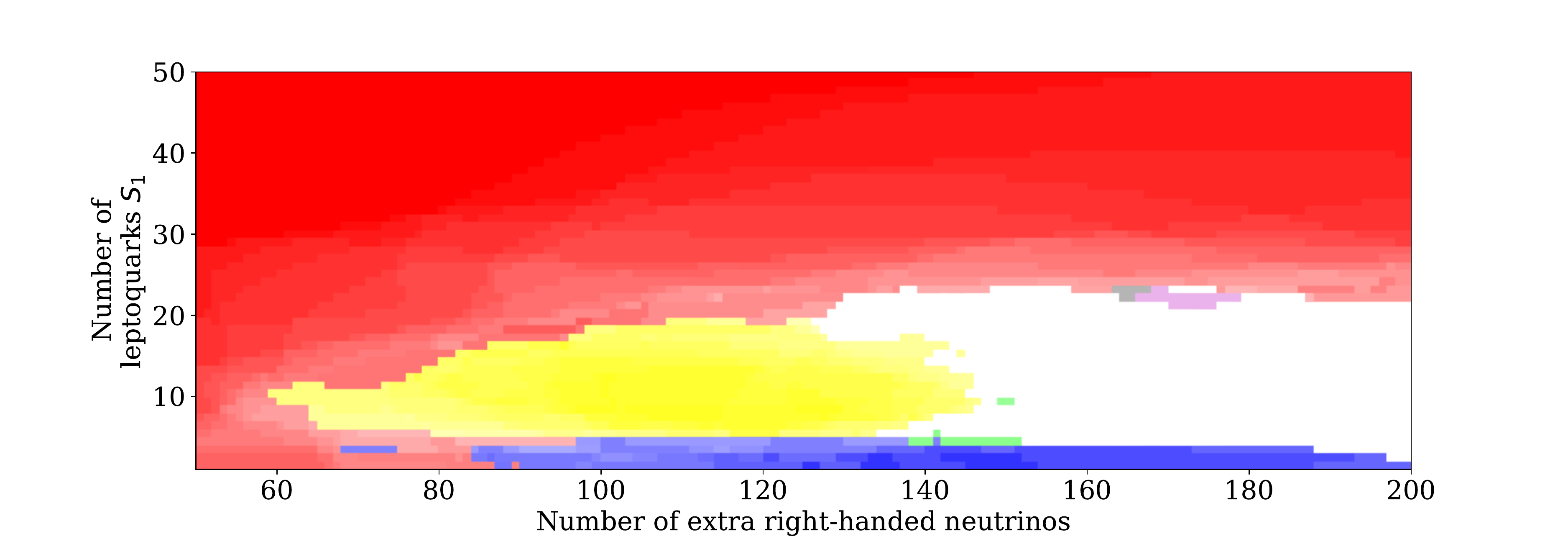}\end{minipage}\\[-2mm]
(d) & \begin{minipage}{\textwidth}\includegraphics[width=\linewidth]{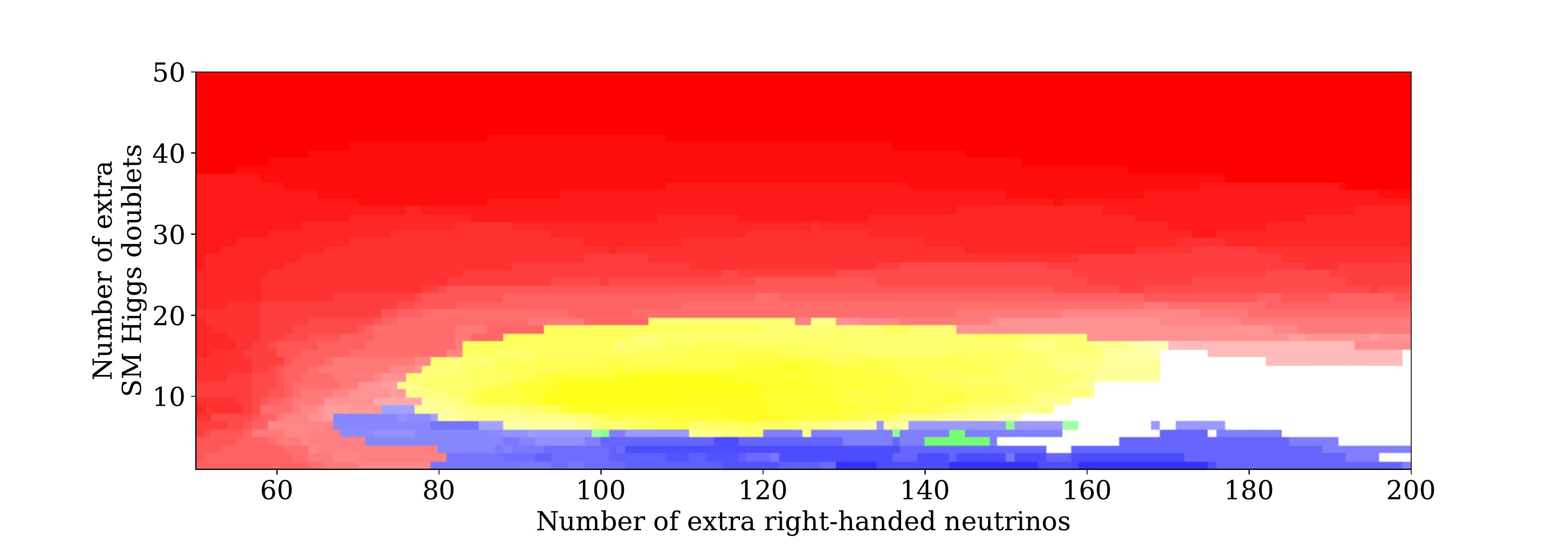}\end{minipage}\\[-2mm]
\multicolumn{2}{c}{\includegraphics[width=\linewidth]{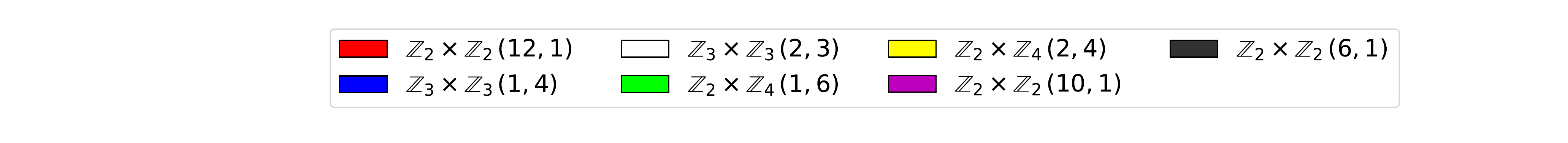}}
\end{tabular}%
\end{center}
\vspace{-1.1cm}
\caption{Predictions of orbifold geometries for SM-like spectra without exotics, except for 110 
scalar flavons $(\rep1, \rep1)_{0}$, and additional particles as indicated by the labels of the 
axes. The transparency of each pixel indicates the accuracy of the respective prediction:
the less transparent the color, the more accurate the prediction.}
\label{fig:predictions1}
\end{figure}

By comparing to our 170,219 explicitly constructed SM-like models, we see that especially the 
orbifold geometries \Z2\x\Z2 (12,1) and \Z3\x\Z3 (1,4) are able to produce the SM spectrum with the 
least number of additional particles compared to all other orbifold geometries. In addition, we 
predict the orbifold origin of SM-like models with 110 SM scalar singlets (roughly at the average 
number for all orbifold geometries, see table~\ref{tab:AN-es}) and with some scalar 
leptoquarks $S_1$, $\bar{S}_1$, and $\tilde{R}_2$, but no further exotics. The results are 
illustrated in figs.~\ref{fig:predictions1}~(a)--(c). Finally, we predict the orbifold origin of a 
SM-like model with no other exotics except for a (large) number of additional Higgs doublets, 
see fig.~\ref{fig:predictions1}~(d). Recall that we have found correlations between the number of 
leptoquarks $\bar{S}_1$ and $\tilde{R}_2$, and between the number of leptoquarks $S_1$ and the 
number of additional Higgs doublets, see the respective plots in fig.~\ref{fig:correlations}. 
Therefore, it is also clear why the predictions in figs.~\ref{fig:predictions1}~(a) 
and~\ref{fig:predictions1}~(b) are very similar, as well as the predictions in 
figs.~\ref{fig:predictions1}~(c) and~\ref{fig:predictions1}~(d).

\clearpage

\section{Benchmark SM-like models}
\label{sec:benchmark}

In this section, we discuss some details of three benchmark SM-like string models. Two of them are 
characteristic {\it almost SM} orbifold models: one without exotic scalars and one without exotic 
fermions. The third model is a SM-like model including a small number of right-handed neutrinos and 
SM-singlet scalars. These models arise from different \Z{N}\x\Z{M} orbifold geometries. Hence, they 
are defined by these geometries and their gauge embedding in terms of the 16D shift vectors 
$V_1,V_2$ and Wilson lines $W_\alpha$, $\alpha=1,\ldots,6$.

\begin{table}[t!]
\begin{center}
{\footnotesize
\begin{tabular}{|rlc|c|rlc|c|rlc|}
  \cline{1-3}\cline{5-7}\cline{9-11}
  \multicolumn{3}{|c|}{Model 1} && \multicolumn{3}{c|}{Model 2} && \multicolumn{3}{c|}{Model 3}\\
  \cline{1-3}\cline{5-7}\cline{9-11}
  \#  & fermionic irrep & label && \#  & fermionic irrep & label && \#  & fermionic irrep & label \\
  \cline{1-3}\cline{5-7}\cline{9-11}
  5 & $(\rep1, \rep2)_{\nicefrac{-1}{2}}$  & $\ell_i$        &&   3 & $(\rep1, \rep2)_{\nicefrac{-1}{2}}$   & $\ell_i$      &&
                                                                       3 & $(\rep1, \rep2)_{\nicefrac{-1}{2}}$   & $\ell_i$ \\
  2 & $(\rep1, \rep2)_{\nicefrac{1}{2}}$   & $\bar{\ell}_i$  &&     &                                       &               && 
                                                                       3 &  $(\rep1, \rep1)_{-1}$                & $e_i$    \\ 
  3 & $(\rep1, \rep1)_{1}$                 & $\bar{e}_i$     &&   3 & $(\rep1, \rep1)_{1}$                  & $\bar{e}_i$   &&
                                                                       6 &  $(\rep1, \rep1)_{1}$                 & $\bar{e}_i$ \\ 
  3 & $(\rep3, \rep2)_{\nicefrac{1}{6}}$   & $q_i$           &&   3 & $(\rep3, \rep2)_{\nicefrac{1}{6}}$    & $q_i$         && 
                                                                       3 & $(\rep3, \rep2)_{\nicefrac{1}{6}}$    & $q_i$   \\				
  3 & $(\crep3, \rep1)_{\nicefrac{-2}{3}}$ & $\bar{u}_i$     &&   3 & $(\crep3, \rep1)_{\nicefrac{-2}{3}}$  & $\bar{u}_i$   && 
                                                                       6 & $(\crep3, \rep1)_{\nicefrac{-2}{3}}$  & $\bar{u}_i$\\
    &                                      &                 &&     &                                       &               &&                                                                       
                                                                       3 & $(\rep3, \rep1)_{\nicefrac{2}{3}}$    & $u_i$\\
  7 & $(\crep3, \rep1)_{\nicefrac{1}{3}}$  & $\bar{d}_i$     &&   3 & $(\crep3, \rep1)_{\nicefrac{1}{3}}$   & $\bar{d}_i$   && 
                                                                       5 & $(\crep3, \rep1)_{\nicefrac{1}{3}}$   & $\bar{d}_i$ \\
  4 & $(\rep3, \rep1)_{\nicefrac{-1}{3}}$  & $d_i$           &&     &                                       &              && 
                                                                       2 & $(\rep3, \rep1)_{\nicefrac{-1}{3}}$   & $d_i$ \\
111 & $(\rep1, \rep1)_{0}$                 & $\bar{\nu}_i$   && 119 & $(\rep1, \rep1)_{0}$                  & $\bar{\nu}_i$ && 
                                                                      67 & $(\rep1, \rep1)_{0}$                  & $\bar{\nu}_i$\\
  \cline{1-3}\cline{5-7}\cline{9-11}
  \cline{1-3}\cline{5-7}\cline{9-11}
  \#  & scalar irrep & label && \#  & scalar irrep & label && \#  & scalar irrep & label\\
  \cline{1-3}\cline{5-7}\cline{9-11}
  6 & $(\rep1, \rep2)_{\nicefrac{1}{2}}$  & $\phi_i$         &&   9 & $(\rep1, \rep2)_{\nicefrac{1}{2}}$    & $\phi_i$ &&
                                                                       6 & $(\rep1, \rep2)_{\nicefrac{1}{2}}$  & $\phi_i$\\
 76 & $(\rep1, \rep1)_{0}$                & $s_i$            &&  30 & $(\bs1, \bs1)_{0}$                    & $s_i$    &&
                                                                      24 & $(\rep1, \rep1)_{0}$                & $s_i$ \\
    &                                     &                  &&   9 & $(\crep3, \rep1)_{\nicefrac{1}{3}}$   & $S_{1,i}$ &&
                                                                      12 & $(\crep3,\rep1)_{\nicefrac{1}{3}}$  & $S_{1,i}$\\  
 \cline{1-3}\cline{5-7}\cline{9-11}
\end{tabular}
\caption{Massless spectra of three promising SM-like models. The representations are written with 
respect to $\mathcal{G}_{\mathrm{SM}}=\SU3_C\x\SU2_L\x\U1_Y$, where the hypercharges are displayed 
as subscripts. All fermions are left-handed. Model 1 is an {\it almost SM} based on the 
\Z2\x\Z4 (2,4) orbifold geometry; model 2 is an {\it almost SM} based on the \Z3\x\Z3 (1,4) 
orbifold geometry; and model 3 is an almost SM string model arising from the \Z2\x\Z2 (12,1) 
geometry. Besides the numbers of Higgs doublets $\phi$, right-handed neutrinos $\bar\nu$ and 
flavons $s$, the models differ by the extra ``hidden'' non-Abelian gauge factor which acts as a 
continuous gauge flavor symmetry.}
\label{tab:spectraSM}
}
\end{center}
\end{table}

Our benchmark models are defined as follows:
\begin{itemize}
\item {\bf Model 1.} {\it Almost SM} model based on the orbifold geometry \Z2\x\Z4 (2,4)
and its gauge embedding given by the shift vectors
\begin{subequations}
\begin{eqnarray}
V_1 & = & \tfrac{1}{2}\left(0, 0, 0, 0, 0, 1,1, 4, 0, 0, 0, 0, 1,1,1,1\right)\;,\\
V_2 & = & \tfrac{1}{8}\left(-3, -1, -1, -1, 3, 1,9, -3, -5, -1,1, 1, -1, -1,1, 5\right)\;,
\end{eqnarray}
\end{subequations}
and the Wilson lines (with $W_1=W_2=0$)
\begin{subequations}
\begin{eqnarray}
W_3=W_4=W_6 & = & \tfrac12\left(0, 0, 0, 0, 0, 0, 0, 0, -1, 0 -3, 0 -3, 2, -2, 3\right)\;,\\
W_5         & = & \tfrac14\left(-7, -1, -1, 3, 7, -3, 9, -7, -7, 5, -5, 9, -3, 7, 7, 3\right)\;.
\end{eqnarray}
\end{subequations}
The resulting 4D gauge group is given by
\begin{equation}
{\mathcal{G}}_{4D} ~=~ \SU3_C\x \SU2_L \x \U1_Y \x \mathcal{G}_{\text{hidden}} \x \U1'^8,
\end{equation}
where ${\mathcal{G}}_{\text{hidden}} = \SU3\x\SU2^2$ is the hidden gauge group and one of 
the $\U1'$ is anomalous. The SM gauge quantum numbers of the massless spectrum are presented 
in table~\ref{tab:spectraSM}, where we display explicitly the pairs of vector-like exotic 
fermions. Note that there are three vector-like pairs of extra lepton doublets and four
pairs of exotic down-type quarks, which can develop large masses when some of the 76 
flavons attain VEVs. This model exhibits a total of six Higgs doublets and 111 right-handed
neutrinos.

\item {\bf Model 2.} {\it Almost SM} model based on the orbifold geometry \Z3\x\Z3 (1,4)
and its gauge embedding given by the shift vectors
\begin{subequations}
\begin{eqnarray}
V_1 & = & \tfrac{1}{6}\left(-9, -1, -1, -1, -1, 1, 1, 3, -1, 1, 1, 1, 1, 1, 1, 7\right)\;,\\
V_2 & = & \tfrac{1}{6}\left( 9, -11, -3, 1, 1, -5, 3, -11, -1, -11, -1, -1, -1, 1, 3, 3\right)\;,
\end{eqnarray}
\end{subequations}
and the Wilson lines (with $W_\alpha=W_1$ for all $\alpha$)
\begin{subequations}
\begin{eqnarray}
W_1 & = & \tfrac{1}{6}\left( -1, 5, -9, -1, 3, -9, -9, -11, -11, -7, -1, 1, 5, -1, -11, 13\right)\;.
\end{eqnarray}
\end{subequations}
The 4D gauge group reads
\begin{equation}
\mathcal{G}_{4D} ~=~ \SU3_C\x\SU2_L\x\U1_Y \x\mathcal{G}_{\text{hidden}}\times \U1'^9\,,
\end{equation}
where $\mathcal{G}_{\text{hidden}} = \SU2^3$ is the hidden gauge group and one of the $\U1'$ is 
anomalous. The SM gauge quantum numbers of the massless spectrum are shown in 
table~\ref{tab:spectraSM}. As exotics, this model includes nine $S_1$ leptoquarks and eight 
additional Higgs doublets. Clearly, including the standard Higgs doublet, these fields build 
$\rep5$-plets of local \SU5 GUTs in higher dimensions. In addition, we observe a large set of 
right-handed neutrinos and 30 scalar singlets.

\item {\bf Model 3.} SM-like string model based on the orbifold geometry \Z2\x\Z2 (12,1)
and its gauge embedding given by the shift vectors
\begin{subequations}
\begin{eqnarray}
V_1 & = & \tfrac{1}{2}\left(0, 0, 0, 0, 0, 0, 0, 8, -2, 0, 0, 0, 0, 0, 1, 1\right)\;,\\
V_2 & = & \tfrac{1}{4}\left(-1, -1, 1, 1, 1, 1, 19, 7, -5, -1, -1, -1, -1, 5, -7, -1\right)\;,
\end{eqnarray}
\end{subequations}
and the Wilson lines (with $W_1=W_2=W_3$ and $W_4=W_5=W_6$)
\begin{subequations}
\begin{eqnarray}
W_1 & = & \tfrac{1}{8}\left( 1, 13, -1, -1, -1, 15, 9, 21, 7, -1, -1, 1, 3, -11, 7, 15\right)\;,\\
W_4 & = & \tfrac{1}{4}\left(6, 2, -1, 0, 1, -6, -3, 1, 0, 0, 0, 0, 0, 0, 0, 0\right)\;.
\end{eqnarray}
\end{subequations}
The 4D gauge group is given by
\begin{equation}
{\mathcal{G}}_{4D} ~=~ \SU3_C\x \SU2_L \x \U1_Y \x \mathcal{G}_{\text{hidden}} \x \U1'^{10}\,,
\end{equation}
where ${\mathcal{G}}_{\text{hidden}} = \SU2^2$ is the hidden gauge group and one of the $\U1'$ is 
anomalous. The SM quantum numbers of the matter spectrum of this model are displayed in 
table~\ref{tab:spectraSM}. We observe that this model yields the smallest number of SM singlets 
among the spectra of the benchmark models. However, there is a large number of (pairs of) 
vector-like exotic fermions and $S_1$ leptoquarks. As in many other SM-like models, there are six 
Higgs doublets.
\end{itemize}

\section{Conclusions and outlook}
\label{sec:conclusions}

In this work we have performed the most extensive search for SM-like models from orbifold 
compactification of the non-SUSY heterotic string $\SO{16}\x\SO{16}$. We inspected their massless 
spectra looking for the SM-like models whose spectra best resemble the one of the SM, and for 
useful patterns that may guide us to find the SM from string theory. 

Using a non-SUSY extension of the \texttt{orbifolder} and considering all 138 orbifolds classified 
in ref.~\cite{Fischer:2012qj}, we find 170,219 SM-like models distributed among 104 orbifold 
geometries, as presented in table~\ref{tab:allmodels}. Orbifolds with point groups \Z2\x\Z4 and 
\Z8-II produce the majority of the models with 147,996 out of 169,177 in \Z{N}\x\Z{M} orbifolds and 
423 out of 1,042 in \Z{N} orbifolds. These models include the SM gauge group, three generations of 
SM fermions, including three right-handed neutrinos, at least one Higgs doublet, a number of 
SM-singlet scalars or fermions, and a few vector-like exotic fermions and exotic scalars. We 
classify all (52) possible exotic representations (where some of them behave as leptoquarks), and 
the number (1--55) of Higgs doublets that can appear in SM-like string models. Our results, 
summarized in the tables of appendix~\ref{app:tables-exh}, indicate that only certain types of 
exotic representations appear in these constructions and they are not arbitrary. In particular, 
they build generically representations of \SU5 (local) GUTs at the singularities of the orbifold in 
extra dimensions. Further, most exotic scalars transform either as extra Higgs doublets or as 
$S_1$, $\bar S_1$ or $\tilde R_2$ scalar leptoquarks, see ref.~\cite{Dorsner:2016wpm} for notation 
and refs.~\cite{Cai:2017wry,Choi:2018stw,Becirevic:2018afm,Dorsner:2019itg,Lee:2021jdr} for their 
phenomenology.

We explore the massless spectra of our SM-like models in order to identify special SM-like string 
models called here {\it almost SM} models and exhibiting either i) no exotic fermions or ii) no 
exotic scalars, except for SM singlets that may play the role of right-handed neutrinos in the 
fermionic sector and flavons or dark matter candidates in the scalar sector. The details of these 
models are discussed in section~\ref{sec:almostSM} and summarized in the tables of our 
appendix~\ref{app:ASMtables}.

In section~\ref{sec:ML}, we apply machine learning techniques to our dataset of 170,219 SM-like 
string models. Following ref.~\cite{Parr:2020oar}, we train a neural network such that it predicts, 
based on a requested particle spectrum, the orbifold geometry that most likely can host the 
corresponding SM-like string model. Our analysis shows that the underlying orbifold geometry leaves 
a distinct imprint on the matter spectrum of the resulting SM-like string model. We are thus able 
to predict the phenomenologically most promising orbifold geometries to be \Z2\x\Z2 (12,1), 
\Z3\x\Z3 (1,4), \Z2\x\Z4 (2,4) and \Z3\x\Z3 (2,3), see figures~\ref{fig:predictions2} and~\ref{fig:predictions1}. 
Note that we make the list of all particle spectra available and invite the community to  
consider them in their studies. This information can be found in our website~\cite{website:2021}.
Our data includes i) the files that contain shifts and Wilson lines 
of all 170,219 SM-like string models, ii) files that contain the {\it almost SM} models, and 
iii) the complete list of exotics for all 170,219 SM-like models.

To illustrate the qualities of our models, we present in section~\ref{sec:benchmark} three
special models, two {\it almost SM} models and one SM-like model with a reduced number of SM singlets.
They correspond to a sample of the models that arise from the three most promising orbifold 
geometries identified by using machine learning techniques.

One task beyond this work is the detailed study of the phenomenology of our SM-like string models. 
With this purpose, one should first construct the interaction terms $\mathcal{L}_\mathrm{int.}$ 
that give rise to the couplings among the different SM fields and the (scalar and fermionic) 
exotics. From this, one could obtain constraints on the parameters of the couplings that may lead 
to rapid proton decay, that could explain the $g_\mu-2$ discrepancy via leptoquarks, or that could 
provide admissible scenarios of multi-Higgs portals to dark matter, among other scenarios. Some of 
these phenomenological questions shall be studied elsewhere.

Another interesting endeavor in the non-SUSY heterotic string compactified on orbifolds is the study of 
the emerging eclectic flavor scheme, which is the natural nontrivial combination of traditional and 
modular flavor symmetries appearing in string orbifolds~\cite{Nilles:2020kgo,Nilles:2020tdp,Baur:2020jwc}. 
The eclectic picture has been studied only in the supersymmetric context, i.e.\ in the case of 
orbifold compactifications of the \E8\x\E8 heterotic string. As modular and traditional flavor 
symmetries originate from the outer automorphisms of the Narain space group, which are independent 
of the presence of supersymmetry, extending the discussion to the non-SUSY case should be feasible. 
This would, on the one hand, provide an understanding of modular flavor symmetries without 
SUSY, and on the other, complete the classification of all possible flavor symmetries emerging 
from orbifold compactifications~\cite{Kobayashi:2006wq,Olguin-Trejo:2018wpw}.

The final goal of the construction of non-supersymmetric string constructions is to arrive at
a phenomenologically viable model. This requires to address the questions of potential instabilities 
beyond the perturbatively tachyon-free spectra presented here and the potentially large cosmological 
constant, as discussed in e.g.\ 
refs.~\cite{Satoh:2015nlc,Abel:2015oxa,Abel:2017vos,GrootNibbelink:2017luf,Aoyama:2020aaw,Aoyama:2021kqa}.
We postpone the study of these challenges for future works.

\section*{Acknowledgments}

P.V. is supported by the Deutsche Forschungsgemeinschaft (SFB1258). We would like to thank
J.~Armando Arroyo, Esa\'u Cervantes, Yessenia Olgu\'in-Trejo and Omar P\'erez-Figueroa for useful
discussions.

\begin{appendix}

\begin{landscape}

\section{Exotic matter and Higgs doublets of stringy SM-like models} 
\label{app:tables-exh}

\begin{table}[h!]
\begin{center}
\resizebox{1.308\textwidth}{!}{
\begin{tabular}{|l|r|r|r|r|r|r|r|r|r|r|r|r|r|r|r|}
\hline
                      & $\Z3$ & $\Z4$ & $\Z6$-I & $\Z6$-II & $\Z8$-I & $\Z8$-II & $\Z{12}$-II & $\Z2\times\Z2$  & $\Z2\times\Z4$ & $\Z2\times\Z6$-I & $\Z2\times\Z6$-II & $\Z3\times\Z3$ & $\Z3\times\Z6$ & $\Z4\times\Z4$ & $\Z6\times\Z6$ \\ \hline
\# SM                 & 155 & 30 & 64 & 260 & 8 & 423 & 102 & 234 & 147,996 & 606 & 457 & 14,891 & 147 & 4,834 &  12  \\\hline\hline
$(\rep3,\rep2)_{\nicefrac{1}{6}}$ & 0\%     & 10.00\% & 45.31\% & 40.00\% & 100\%   & 21.51\% & 2.94\%  & 62.39\% & 32.55\% & 48.84\%  & 64.11\% & 0.56\%  & 58.50\% & 22.34\% & 16.67\% \\
$(\crep3,\rep1)_{\nicefrac{-2}{3}}$ & 0\%     & 3.33\%  & 0\%     & 23.85\% & 0\%     & 20.80\% & 2.94\%  & 57.69\% & 33.04\% & 50.33\%  & 23.19\% & 0.46\%  & 67.35\% & 20.15\% & 66.67\% \\
$(\crep3,\rep1)_{\nicefrac{1}{3}}$  & 83.87\% & 100\%   & 100\%   & 94.62\% & 100\%   & 96.93\% & 86.27\% & 99.15\% & 98.90\% & 96.20\%  & 100\%   & 96.56\% & 100\%   & 99.44\% & 100\%   \\
$(\rep1,\rep2)_{\nicefrac{-1}{2}}$           & 62.58\% & 96.67\% & 90.63\% & 96.54\% & 100\%   & 95.98\% & 86.27\% & 97.01\% & 98.81\% & 96.70\%  & 97.37\% & 94.92\% & 94.56\% & 99.13\% & 100\%   \\
$(\rep1,\rep1)_{1}$               & 0\%     & 3.33\%  & 0\%     & 22.69\% & 0\%     & 20.80\% & 2.94\%  & 57.69\% & 33.04\% & 50.33\%  & 23.19\% & 0.61\%  & 68.03\% & 20.02\% & 66.67\% \\ \hline
$(\rep1,\rep1)_{0}$               & 100\%   & 100\%   & 100\%   & 100\%   & 100\%   & 100\%   & 100\%   & 100\%   & 100\%   & 100\%    & 100\%   & 100\%   & 100\%   & 100\%   & 100\%   \\ \hline
$(\rep3,\rep2)_{\nicefrac{-1}{3}}$            & 0\%     & 0\%     & 0\%     & 0\%     & 0\%     & 0\%     & 0\%     & 0\%     & 2.16\%  & 0.17\%   & 0\%     & 0\%     & 1.36\%  & 0.04\%  & 0\%     \\
$(\rep3,\rep2)_{\nicefrac{-1}{6}}$            & 0\%     & 0\%     & 0\%     & 0\%     & 0\%     & 0\%     & 0\%     & 0\%     & 0\%     & 0\%      & 0\%     & 2.41\%  & 0\%     & 0\%     & 0\%     \\
$(\rep3,\rep2)_{\nicefrac{-5}{6}}$           & 49.68\% & 0\%     & 0\%     & 20.38\% & 0\%     & 2.60\%  & 1.96\%  & 0\%     & 0.53\%  & 8.91\%   & 8.75\%  & 0.36\%  & 42.18\% & 5.73\%  & 33.33\% \\  \hline
$(\rep3,\rep1)_{0}$               & 33.55\% & 0\%     & 0\%     & 2.31\%  & 0\%     & 0\%     & 0\%     & 0\%     & 0\%     & 0\%      & 0\%     & 43.37\% & 2.04\%  & 0\%     & 0\%     \\
$(\crep3,\rep1)_{\nicefrac{2}{3}}$   & 0\%     & 0\%     & 0\%     & 0\%     & 0\%     & 0\%     & 0\%     & 0\%     & 0\%     & 0\%      & 0\%     & 1.59\%  & 0\%     & 0\%     & 0\%     \\
$(\crep3,\rep1)_{\nicefrac{-1}{3}}$  & 10.97\% & 0\%     & 0\%     & 0.38\%  & 0\%     & 0\%     & 0\%     & 0\%     & 0\%     & 0\%      & 0\%     & 22.68\% & 1.36\%  & 0\%     & 0\%     \\
$(\rep3,\rep1)_{\nicefrac{-5}{6}}$            & 0\%     & 0\%     & 0\%     & 0\%     & 0\%     & 0\%     & 0\%     & 0\%     & 0\%     & 0\%      & 0\%     & 0\%     & 1.36\%  & 0.08\%  & 0\%     \\
$(\rep3,\rep1)_{\nicefrac{1}{6}}$             & 49.03\% & 66.67\% & 84.38\% & 88.08\% & 25.00\% & 65.96\% & 56.86\% & 73.08\% & 68.88\% & 65.02\%  & 100\%   & 0.38\%  & 89.12\% & 76.89\% & 100\%   \\
$(\rep3,\rep1)_{\nicefrac{-1}{12}}$           & 0\%     & 0\%     & 0\%     & 0\%     & 0\%     & 0\%     & 0\%     & 0\%     & 0\%     & 0\%      & 0\%     & 0\%     & 0\%     & 0.19\%  & 0\%     \\ \hline
$(\rep1,\rep2)_{0}$               & 50.32\% & 100\%   & 100\%   & 95.38\% & 100\%   & 98.58\% & 96.08\% & 71.37\% & 76.95\% & 99.83\%  & 100\%   & 0.39\%  & 94.56\% & 99.26\% & 100\%   \\
$(\rep1,\rep2)_{\nicefrac{1}{3}}$             & 0\%     & 0\%     & 0\%     & 0\%     & 0\%     & 0\%     & 0\%     & 0\%     & 0\%     & 0\%      & 0\%     & 0.01\%  & 0\%     & 0\%     & 0\%     \\
$(\rep1,\rep2)_{\nicefrac{1}{4}}$             & 0\%     & 0\%     & 0\%     & 0\%     & 0\%     & 0\%     & 0\%     & 0\%     & 0\%     & 0\%      & 0\%     & 0\%     & 0\%     & 0.17\%  & 0\%     \\
$(\rep1,\rep2)_{\nicefrac{1}{6}}$             & 50.32\% & 0\%     & 0\%     & 2.69\%  & 0\%     & 0\%     & 0\%     & 0\%     & 0\%     & 0\%      & 0\%     & 58.48\% & 2.04\%  & 0\%     & 0\%     \\
$(\rep1,\rep2)_{\nicefrac{5}{6}}$             & 0\%     & 0\%     & 0\%     & 0\%     & 0\%     & 0\%     & 0\%     & 0\%     & 0\%     & 0\%      & 0\%     & 0.01\%  & 0\%     & 0\%     & 0\%     \\ \hline
$(\rep1,\rep1)_{\nicefrac{1}{2}}$             & 50.32\% & 100\%   & 100\%   & 97.69\% & 100\%   & 100\%   & 100\%   & 91.03\% & 81.41\% & 100.00\% & 100\%   & 0.39\%  & 97.96\% & 100\%   & 100\%   \\
$(\rep1,\rep1)_{\nicefrac{1}{3}}$             & 50.32\% & 0\%     & 0\%     & 2.69\%  & 0\%     & 0\%     & 0\%     & 0\%     & 0\%     & 0\%      & 0\%     & 86.48\% & 2.04\%  & 0\%     & 0\%     \\
$(\rep1,\rep1)_{\nicefrac{2}{3}}$             & 50.32\% & 0\%     & 0\%     & 2.69\%  & 0\%     & 0\%     & 0\%     & 0\%     & 0\%     & 0\%      & 0\%     & 55.06\% & 2.04\%  & 0\%     & 0\%     \\
$(\rep1,\rep1)_{\nicefrac{1}{4}}$             & 0\%     & 0\%     & 0\%     & 0\%     & 0\%     & 0\%     & 0\%     & 0\%     & 0\%     & 0\%      & 0\%     & 0\%     & 0\%     & 0.23\%  & 0\%     \\
$(\rep1,\rep1)_{\nicefrac{3}{4}}$             & 0\%     & 0\%     & 0\%     & 0\%     & 0\%     & 0\%     & 0\%     & 0\%     & 0\%     & 0\%      & 0\%     & 0\%     & 0\%     & 0.10\%  & 0\%     \\
$(\rep1,\rep1)_{\nicefrac{1}{6}}$             & 0.65\%  & 0\%     & 0\%     & 0.38\%  & 0\%     & 0\%     & 0\%     & 0\%     & 0\%     & 0\%      & 0\%     & 0.03\%  & 0\%     & 0\%     & 0\%   \\ \hline 
\end{tabular}
}
\caption{Percentages of SM-like models containing the various types of vector-like exotic fermions. 
We provide in the header the Abelian orbifold point groups where SM-like models were found.  
The row \#SM lists the number of SM-like models arising from all orbifold geometries sharing 
the same point group. We count here vector-like pairs of exotic left-chiral fermions, such that each 
row includes a representation and its complex conjugate, e.g.\ $(\rep3,\rep2)_{\nicefrac{1}{6}}$ stands for 
$(\rep3,\rep2)_{\nicefrac{1}{6}} \oplus (\crep3,\rep2)_{\nicefrac{-1}{6}}$. All fermions beyond the 
three SM generations are considered as exotics.}
\label{tab:VLERf}
\end{center}
\end{table}

\end{landscape}


\begin{landscape}
\begin{table}[p]
\begin{center}
\resizebox{1.308\textwidth}{!}{
\begin{tabular}{|l|r|r|r|r|r|r|r|r|r|r|r|r|r|r|r|}
\hline
                      & $\Z3$ & $\Z4$ & $\Z6$-I & $\Z6$-II & $\Z8$-I & $\Z8$-II & $\Z{12}$-II & $\Z2\times\Z2$ & $\Z2\times\Z4$ & $\Z2\times\Z6$-I & $\Z2\times\Z6$-II & $\Z3\times\Z3$ & $\Z3\times\Z6$ & $\Z4\times\Z4$ & $\Z6\times\Z6$ \\ \hline
\# SM                 & 155 & 30 & 64 & 260 & 8 & 423 & 102 & 234 & 147,996 & 606 & 457 & 14,891 & 147 & 4,834 & 12   \\\hline\hline
$(\rep1,\rep2)_{\nicefrac{1}{2}}$             & 100\%   & 100\%   & 79.69\% & 100\%   & 100\% & 100\%   & 100\%   & 100\%   & 100\%   & 100\%   & 100\%   & 100\%   & 100\%   & 100\%   & 100\%   \\  \hline
$(\rep1,\rep1)_{0}$               & 100\%   & 100\%   & 100\%   & 100\%   & 100\% & 100\%   & 100\%   & 100\%   & 100\%   & 100\%   & 100\%   & 100\%   & 100\%   & 100\%   & 100\%   \\  \hline
$(\rep3,\rep2)_{\nicefrac{-1}{3}}$            & 0\%     & 0\%     & 0\%     & 0\%     & 0\%   & 0\%     & 0\%     & 0\%     & 3.50\%  & 4.79\%  & 10.94\% & 0\%     & 4.76\%  & 2.98\%  & 8.33\%  \\
$(\rep3,\rep2)_{\nicefrac{1}{6}}$             & 49.68\% & 53.33\% & 0\%     & 70.00\% & 100\% & 34.99\% & 21.57\% & 41.03\% & 61.65\% & 67.49\% & 22.76\% & 0.73\%  & {51.02}\% & 45.39\% & 75.00\% \\
$(\rep3,\rep2)_{\nicefrac{-1}{6}}$            & 0\%     & 0\%     & 0\%     & 0\%     & 0\%   & 0\%     & 0\%     & 0\%     & 0\%     & 0\%     & 0\%     & 1.83\%  & 0\%     & 0\%     & 0\%     \\  \hline
$(\rep3,\rep1)_{0}$               & 32.26\% & 0\%     & 0\%     & 1.92\%  & 0\%   & 0\%     & 0\%     & 0\%     & 0\%     & 0\%     & 0\%     & 21.61\% & 0.68\%  & 0\%     & 0\%     \\
$(\crep3,\rep1)_{\nicefrac{2}{3}}$  & 0\%     & 0\%     & 0\%     & 0\%     & 0\%   & 0\%     & 0\%     & 0\%     & 0\%     & 0\%     & 0\%     & 5.68\%  & 0\%     & 0\%     & 0\%     \\
$(\crep3,\rep1)_{\nicefrac{-2}{3}}$ & 49.68\% & 30.00\% & 0\%     & 60.00\% & 0\%   & 40.43\% & 25.49\% & {41.45}\% & 61.21\% & 42.08\% & 31.29\% & 0.73\%  & {50.34}\% & 40.75\% & 100\%   \\
$(\crep3,\rep1)_{\nicefrac{1}{3}}$  & 100\%   & 100\%   & 100\%   & 100\%   & 100\% & 100\%   & 100\%   & 100\%   & 98.58\% & 100\%   & 100\%   & 99.93\% & 100\%   & 100\%   & 100\%   \\
$(\crep3,\rep1)_{\nicefrac{-1}{3}}$ & 5.81\%  & 0\%     & 0\%     & 0.38\%  & 0\%   & 0\%     & 0\%     & 0\%     & 0\%     & 0\%     & 0\%     & 12.06\% & 0\%     & 0\%     & 0\%     \\
$(\rep3,\rep1)_{\nicefrac{1}{6}}$             & 0\%     & 70.00\% & 100\%   & 92.69\% & 100\% & 79.67\% & 78.43\% & {71.37}\% & 70.14\% & 72.44\% & 98.91\% & 0\%     & {81.63}\% & 90.13\% & 100\%   \\
$(\rep3,\rep1)_{\nicefrac{-1}{6}}$            & 0\%     & 0\%     & 0\%     & 0.38\%  & 0\%   & 0\%     & 0\%     & 0\%     & 0\%     & 0\%     & 0\%     & 0\%     & 0\%     & 0\%     & 0\%     \\
$(\rep3,\rep1)_{\nicefrac{-5}{6}}$            & 0\%     & 0\%     & 0\%     & 0\%     & 0\%   & 0\%     & 0\%     & 0\%     & 0\%     & 0\%     & 0\%     & 0\%     & {3.40}\%  & 0.17\%  & 41.67\% \\
$(\rep3,\rep1)_{\nicefrac{-1}{12}}$           & 0\%     & 0\%     & 0\%     & 0\%     & 0\%   & 0\%     & 0\%     & 0\%     & 0\%     & 0\%     & 0\%     & 0\%     & 0\%     & 0.10\%  & 0\%     \\
$(\rep3,\rep1)_{\nicefrac{5}{12}}$            & 0\%     & 0\%     & 0\%     & 0\%     & 0\%   & 0\%     & 0\%     & 0\%     & 0\%     & 0\%     & 0\%     & 0\%     & 0\%     & 0.06\%  & 0\%     \\ \hline
$(\rep1,\rep2)_{0}$               & 0\%     & 100\%   & 100\%   & 87.31\% & 100\% & 98.35\% & 98.04\% & {74.36}\% & 77.04\% & 97.19\% & 98.91\% & 0\%     & {82.31}\% & 99.19\% & 100\%   \\
$(\rep1,\rep2)_{\nicefrac{1}{6}}$             & 31.61\% & 0\%     & 0\%     & 2.31\%  & 0\%   & 0\%     & 0\%     & 0\%     & 0\%     & 0\%     & 0\%     & 32.60\% & {2.04}\%  & 0\%     & 0\%     \\
$(\rep1,\rep2)_{\nicefrac{5}{6}}$             & 0\%     & 0\%     & 0\%     & 0\%     & 0\%   & 0\%     & 0\%     & 0\%     & 0\%     & 0\%     & 0\%     & 1.13\%  & 0\%     & 0\%     & 0\%     \\
$(\rep1,\rep2)_{\nicefrac{1}{4}}$             & 0\%     & 0\%     & 0\%     & 0\%     & 0\%   & 0\%     & 0\%     & 0\%     & 0\%     & 0\%     & 0\%     & 0\%     & 0\%     & 0.19\%  & 0\%     \\  \hline
$(\rep1,\rep1)_{1}$               & 51.61\% & 30.00\% & 0\%     & 59.23\% & 0\%   & 40.43\% & 25.49\% & {41.45}\% & 61.21\% & 42.08\% & 31.29\% & 0.73\%  & {50.34}\% & 40.73\% & 100\%   \\
$(\rep1,\rep1)_{\nicefrac{1}{2}}$             & 0\%     & 100\%   & 100\%   & 97.69\% & 100\% & 100\%   & 100\%   & {91.45}\% & 81.40\% & 100\%   & 100\%   & 0\%     & {97.96}\% & 100\%   & 100\%   \\
$(\rep1,\rep1)_{\nicefrac{1}{3}}$             & 50.32\% & 0\%     & 0\%     & 2.69\%  & 0\%   & 0\%     & 0\%     & 0\%     & 0\%     & 0\%     & 0\%     & 78.15\% & {2.04}\%  & 0\%     & 0\%     \\
$(\rep1,\rep1)_{\nicefrac{2}{3}}$             & 29.68\% & 0\%     & 0\%     & 2.69\%  & 0\%   & 0\%     & 0\%     & 0\%     & 0\%     & 0\%     & 0\%     & 39.43\% & {2.04}\%  & 0\%     & 0\%     \\
$(\rep1,\rep1)_{\nicefrac{1}{4}}$             & 0\%     & 0\%     & 0\%     & 0\%     & 0\%   & 0\%     & 0\%     & 0\%     & 0\%     & 0\%     & 0\%     & 0\%     & 0\%     & 0.23\%  & 0\%     \\
$(\rep1,\rep1)_{\nicefrac{3}{4}}$             & 0\%     & 0\%     & 0\%     & 0\%     & 0\%   & 0\%     & 0\%     & 0\%     & 0\%     & 0\%     & 0\%     & 0\%     & 0\%     & 0.12\%  & 0\%     \\
$(\rep1,\rep1)_{\nicefrac{1}{6}}$             & 0\%     & 0\%     & 0\%     & 0.38\%  & 0\%   & 0\%     & 0\%     & 0\%     & 0\%     & 0\%     & 0\%     & 0\%     & 0\%     & 0\%     & 0\% \\ \hline   
\end{tabular}
}
\caption{Percentages of SM-like models containing the various types of exotic scalars. 
We provide in the header the Abelian orbifold point groups where SM-like models were found. 
The row \#SM lists the number of SM-like models arising from all orbifold geometries sharing 
the same point group. Higgs doublets beyond the SM Higgs are considered as exotics. Since we 
consider complex scalars, we identify a representation and its complex conjugate, e.g.
$(\rep3,\rep1)_{\nicefrac{1}{6}} \equiv (\crep3,\rep1)_{\nicefrac{-1}{6}}$.}
\label{tab:VLERs}
\end{center}
\end{table}
\end{landscape}

\begin{landscape}
\begin{table}[hp]
\begin{center}
\resizebox{1.308\textwidth}{!}{
\begin{tabular}{|l|r|r|r|r|r|r|r|r|r|r|r|r|r|r|r|}
\hline
                      & $\Z3$ & $\Z4$ & $\Z6$-I & $\Z6$-II & $\Z8$-I & $\Z8$-II & $\Z{12}$-II & $\Z2\times\Z2$  & $\Z2\times\Z4$ & $\Z2\times\Z6$-I & $\Z2\times\Z6$-II & $\Z3\times\Z3$ & $\Z3\times\Z6$ & $\Z4\times\Z4$ & $\Z6\times\Z6$ \\ \hline
\# SM                 & 155 & 30 & 64 & 260 & 8 & 423 & 102 & 234 & 147,996 & 606 & 457 & 14,891 & 147 & 4,834 &  12  \\\hline\hline
$(\rep3,\rep2)_{\nicefrac{1}{6}}$             & 0.00   & 0.10   & 0.45   & 0.95   & 1.00   & 0.24   & 0.06   & {1.18}   & 0.34   & 0.84   & 0.64   & 0.01   & {1.01}   & 0.23   & 0.25   \\
$(\crep3,\rep1)_{\nicefrac{-2}{3}}$ & 0.00   & 0.03   & 0      & 0.28   & 0      & 0.21   & 0.06   & 1.12   & 0.35   & 0.89   & 0.24   & 0.01   & 0.95   & 0.21   & 0.75   \\
$(\crep3,\rep1)_{\nicefrac{1}{3}}$  & 3.23   & 1.93   & 4.25   & 4.27   & 7.25   & 4.13   & 3.26   & {5.24}   & 6.78   & 7.27   & 7.04   & 3.32   & {8.95}   & 6.91   & 9.00   \\
$(\rep1,\rep2)_{\nicefrac{-1}{2}}$            & 2.50   & 1.67   & 2.80   & 4.26   & 7.00   & 4.13   & 3.30   & {5.32}   & 6.61   & 7.08   & 5.30   & 2.86   & {5.82}   & 6.51   & 3.75   \\
$(\rep1,\rep1)_{1}$               & 0      & 0.03   & 0      & 0.27   & 0      & 0.21   & 0.06   & 1.12   & 0.35   & 0.89   & 0.24   & 0.01   & {0.96}   & 0.21   & 0.75   \\ \hline
$(\rep1,\rep1)_{0}$               & 116.99 & 107.47 & 194.13 & 135.41 & 205.50 & 148.08 & 117.84 & {151.35} & 174.23 & 183.41 & 195.47 & 121.03 & {207.76} & 212.27 & 167.33 \\ \hline
$(\rep3,\rep2)_{\nicefrac{-1}{3}}$            & 0      & 0      & 0      & 0      & 0      & 0      & 0      & 0      & 0.03   & 0.00   & 0      & 0      & 0.03   & 0.00   & 0      \\
$(\rep3,\rep2)_{\nicefrac{-1}{6}}$            & 0      & 0      & 0      & 0      & 0      & 0      & 0      & 0      & 0      & 0      & 0      & 0.04   & 0      & 0      & 0      \\
$(\rep3,\rep2)_{\nicefrac{-5}{6}}$            & 0.50   & 0      & 0      & 0.20   & 0      & 0.03   & 0.02   & 0      & 0.01   & 0.09   & 0.09   & 0.00   & {0.42}   & 0.06   & 0.33   \\  \hline
$(\rep3,\rep1)_{0}$               & 1.21   & 0      & 0      & 0.04   & 0      & 0      & 0      & 0      & 0      & 0      & 0      & 1.44   & 0.04   & 0      & 0      \\
$(\crep3,\rep1)_{\nicefrac{2}{3}}$  & 0      & 0      & 0      & 0      & 0      & 0      & 0      & 0      & 0      & 0      & 0      & 0.02   & 0      & 0      & 0      \\
$(\crep3,\rep1)_{\nicefrac{-1}{3}}$ & 0.33   & 0      & 0      & 0.01   & 0      & 0      & 0      & 0      & 0      & 0      & 0      & 0.52   & 0.03   & 0      & 0      \\
$(\rep3,\rep1)_{\nicefrac{-5}{6}}$            & 0      & 0      & 0      & 0      & 0      & 0      & 0      & 0      & 0      & 0      & 0      & 0      & 0.01   & 0.00   & 0      \\
$(\rep3,\rep1)_{\nicefrac{1}{6}}$             & 2.42   & 0.87   & 2.80   & 4.91   & 1.00   & 2.38   & 0.89   & {1.65}   & 3.14   & 3.31   & 4.29   & 0.03   & 6.97   & 3.63   & 12.42  \\
$(\rep3,\rep1)_{\nicefrac{-1}{12}}$           & 0      & 0      & 0      & 0      & 0      & 0      & 0      & 0      & 0      & 0      & 0      & 0      & 0      & 0.00   & 0      \\  \hline
$(\rep1,\rep2)_{0}$               & 10.85  & 9.80   & 9.06   & 13.68  & 10.50  & 8.90   & 5.41   & {3.26}   & 10.54  & 14.51  & 13.80  & 0.14   & {27.59}  & 11.96  & 30.67  \\
$(\rep1,\rep2)_{\nicefrac{1}{3}}$             & 0      & 0      & 0      & 0      & 0      & 0      & 0      & 0      & 0      & 0      & 0      & 0.00   & 0      & 0      & 0      \\
$(\rep1,\rep2)_{\nicefrac{1}{4}}$             & 0      & 0      & 0      & 0      & 0      & 0      & 0      & 0      & 0      & 0      & 0      & 0      & 0      & 0.01   & 0      \\
$(\rep1,\rep2)_{\nicefrac{1}{6}}$             & 3.43   & 0      & 0      & 0.10   & 0      & 0      & 0      & 0      & 0      & 0      & 0      & 2.34   & 0.16   & 0      & 0      \\
$(\rep1,\rep2)_{\nicefrac{5}{6}}$             & 0      & 0      & 0      & 0      & 0      & 0      & 0      & 0      & 0      & 0      & 0      & 0.00   & 0      & 0      & 0      \\ \hline
$(\rep1,\rep1)_{\nicefrac{1}{2}}$             & 23.45  & 23.67  & 24.14  & 32.43  & 28.00  & 23.83  & 14.86  & {9.31}   & 23.58  & 26.53  & 30.90  & 0.18   & {54.29}  & 28.55  & 63.08  \\
$(\rep1,\rep1)_{\nicefrac{1}{3}}$             & 28.88  & 0      & 0      & 0.76   & 0      & 0      & 0      & 0      & 0      & 0      & 0      & 15.27  & {0.96}   & 0      & 0      \\
$(\rep1,\rep1)_{\nicefrac{2}{3}}$             & 3.25   & 0      & 0      & 0.09   & 0      & 0      & 0      & 0      & 0      & 0      & 0      & 2.28   & 0.10   & 0      & 0      \\
$(\rep1,\rep1)_{\nicefrac{1}{4}}$             & 0      & 0      & 0      & 0      & 0      & 0      & 0      & 0      & 0      & 0      & 0      & 0      & 0      & 0.07   & 0      \\
$(\rep1,\rep1)_{\nicefrac{3}{4}}$             & 0      & 0      & 0      & 0      & 0      & 0      & 0      & 0      & 0      & 0      & 0      & 0      & 0      & 0.00   & 0      \\
$(\rep1,\rep1)_{\nicefrac{1}{6}}$             & 0.33   & 0      & 0      & 0.06   & 0      & 0      & 0      & 0      & 0      & 0      & 0      & 0.00   & 0      & 0      & 0  \\ \hline
\end{tabular}
}
\caption{Average numbers of vector-like exotic fermions for SM-like models. In the first row the orbifold, labeled by its point group, is displayed. The total number of SM-like models including all geometries of a point group is presented in the second row. Hypercharge is normalized such that $(\rep3,\rep2)_{\nicefrac{1}{6}}$ is a left-chiral quark-doublet. We count vector-like pairs, such each row includes a representation and its complex conjugate, e.g. $(\rep3,\rep2)_{\nicefrac{1}{6}}$ stands for $(\rep3,\rep2)_{\nicefrac{1}{6}} \oplus (\crep3,\rep2)_{\nicefrac{-1}{6}}$.}
\label{tab:AN-VLEf}
\end{center}
\end{table}
\end{landscape}

\begin{landscape}
\begin{table}[p]
\begin{center}
\resizebox{1.308\textwidth}{!}{
\begin{tabular}{|l|r|r|r|r|r|r|r|r|r|r|r|r|r|r|r|}
\hline
                      & $\Z3$ & $\Z4$ & $\Z6$-I & $\Z6$-II & $\Z8$-I & $\Z8$-II & $\Z{12}$-II & $\Z2\times\Z2$ & $\Z2\times\Z4$ & $\Z2\times\Z6$-I & $\Z2\times\Z6$-II & $\Z3\times\Z3$ & $\Z3\times\Z6$ & $\Z4\times\Z4$ & $\Z6\times\Z6$ \\ \hline
\# SM                 & 155 & 30 & 64 & 260 & 8 & 423 & 102 & 234 & 147,996 & 606 & 457 & 14,891 & 147 & 4,834 & 12   \\\hline\hline
$(\rep1,\rep2)_{\nicefrac{1}{2}}$             & 7.64  & 5.80  & 3.45  & 9.83   & 14.50  & 7.88   & 5.88  & {10.81}  & 11.27  & 12.97  & 9.32  & 5.44  & {17.34}  & 12.03  & 14.42  \\ \hline
$(\rep1,\rep1)_{0}$               & 81.32 & 94.73 & 75.95 & 106.51 & 139.00 & 115.17 & 85.15 & {105.76} & 136.34 & 127.96 & 96.52 & 41.71 & {122.54} & 172.00 & 163.58 \\  \hline
$(\rep3,\rep2)_{\nicefrac{-1}{3}}$            & 0     & 0     & 0     & 0      & 0      & 0      & 0     & 0      & 0.09   & 0.08   & 0.15  & 0     & 0.08   & 0.06   & 0.08   \\
$(\rep3,\rep2)_{\nicefrac{1}{6}}$             & 1.49  & 1.07  & 0     & 2.09   & 1.00   & 0.50   & 0.25  & {1.33}   & 0.97   & 1.83   & 0.52  & 0.02  & 2.18   & 0.83   & 1.75   \\
$(\rep3,\rep2)_{\nicefrac{-1}{6}}$            & 0     & 0     & 0     & 0      & 0      & 0      & 0     & 0      & 0      & 0      & 0     & 0.06  & 0      & 0      & 0      \\  \hline
$(\rep3,\rep1)_{0}$               & 1.95  & 0     & 0     & 0.04   & 0      & 0      & 0     & 0      & 0      & 0      & 0     & 1.00  & 0.01   & 0      & 0      \\
$(\crep3,\rep1)_{\nicefrac{2}{3}}$  & 0     & 0     & 0     & 0      & 0      & 0      & 0     & 0      & 0      & 0      & 0     & 0.14  & 0      & 0      & 0      \\
$(\crep3,\rep1)_{\nicefrac{-2}{3}}$ & 1.49  & 0.60  & 0     & 1.78   & 0      & 0.57   & 0.29  & {1.34}   & 0.94   & 1.17   & 0.63  & 0.02  & {2.03}   & 0.73   & 3.17   \\
$(\crep3,\rep1)_{\nicefrac{1}{3}}$  & 5.57  & 6.07  & 5.20  & 15.12  & 10.00  & 9.21   & 10.81 & {17.61}  & 10.76  & 16.18  & 8.91  & 6.37  & {11.74}  & 12.33  & 10.83  \\
$(\crep3,\rep1)_{\nicefrac{-1}{3}}$ & 0.25  & 0     & 0     & 0.01   & 0      & 0      & 0     & 0      & 0      & 0      & 0     & 0.44  & 0      & 0      & 0      \\
$(\rep3,\rep1)_{\nicefrac{1}{6}}$             & 0     & 2.27  & 7.44  & 9.10   & 4.50   & 4.38   & 1.95  & {3.27}   & 6.17   & 5.31   & 7.98  & 0     & {7.84}   & 5.92   & 11.33  \\
$(\rep3,\rep1)_{\nicefrac{-1}{6}}$            & 0     & 0     & 0     & 0.01   & 0      & 0      & 0     & 0      & 0      & 0      & 0     & 0     & 0      & 0      & 0      \\
$(\rep3,\rep1)_{\nicefrac{-5}{6}}$            & 0     & 0     & 0     & 0      & 0      & 0      & 0     & 0      & 0      & 0      & 0     & 0     & 0.07   & 0.00   & 0.83   \\
$(\rep3,\rep1)_{\nicefrac{-1}{12}}$           & 0     & 0     & 0     & 0      & 0      & 0      & 0     & 0      & 0      & 0      & 0     & 0     & 0      & 0.00   & 0      \\
$(\rep3,\rep1)_{\nicefrac{5}{12}}$            & 0     & 0     & 0     & 0      & 0      & 0      & 0     & 0      & 0      & 0      & 0     & 0     & 0      & 0.00   & 0      \\  \hline
$(\rep1,\rep2)_{0}$               & 0     & 10.07 & 10.66 & 8.85   & 12.00  & 8.61   & 5.99  & {3.43}   & 10.13  & 10.87  & 11.11 & 0     & {12.49}  & 9.40   & 13.00  \\
$(\rep1,\rep2)_{\nicefrac{1}{6}}$             & 2.39  & 0     & 0     & 0.12   & 0      & 0      & 0     & 0      & 0      & 0      & 0     & 2.15  & 0.14   & 0      & 0      \\
$(\rep1,\rep2)_{\nicefrac{5}{6}}$             & 0     & 0     & 0     & 0      & 0      & 0      & 0     & 0      & 0      & 0      & 0     & 0.03  & 0      & 0      & 0      \\
$(\rep1,\rep2)_{\nicefrac{1}{4}}$             & 0     & 0     & 0     & 0      & 0      & 0      & 0     & 0      & 0      & 0      & 0     & 0     & 0      & 0.01   & 0      \\ \hline
$(\rep1,\rep1)_{1}$               & 1.55  & 0.60  & 0     & 1.78   & 0      & 0.57   & 0.29  & {1.34}   & 0.94   & 1.17   & 0.63  & 0.02  & {2.03}   & 0.73   & 3.17   \\
$(\rep1,\rep1)_{\nicefrac{1}{2}}$             & 0     & 47.33 & 58.13 & 62.74  & 53.50  & 47.04  & 30.79 & {18.91}  & 47.12  & 49.17  & 72.50 & 0     & {75.50}  & 54.33  & 78.67  \\
$(\rep1,\rep1)_{\nicefrac{1}{3}}$             & 32.51 & 0     & 0     & 1.23   & 0      & 0      & 0     & 0      & 0      & 0      & 0     & 17.49 & {1.25}   & 0      & 0      \\
$(\rep1,\rep1)_{\nicefrac{2}{3}}$             & 1.88  & 0     & 0     & 0.13   & 0      & 0      & 0     & 0      & 0      & 0      & 0     & 2.31  & 0.24   & 0      & 0      \\
$(\rep1,\rep1)_{\nicefrac{1}{4}}$             & 0     & 0     & 0     & 0      & 0      & 0      & 0     & 0      & 0      & 0      & 0     & 0     & 0      & 0.14   & 0      \\
$(\rep1,\rep1)_{\nicefrac{3}{4}}$             & 0     & 0     & 0     & 0      & 0      & 0      & 0     & 0      & 0      & 0      & 0     & 0     & 0      & 0.01   & 0      \\
$(\rep1,\rep1)_{\nicefrac{1}{6}}$             & 0     & 0     & 0     & 0.12   & 0      & 0      & 0     & 0      & 0      & 0      & 0     & 0     & 0      & 0      & 0 \\ \hline
\end{tabular}
}
\caption{Average numbers of exotic scalars for SM-like models. In the first row, we 
provide the label of each Abelian point group that we have considered.  In the second row we present the total number of SM-like models in all orbifold geometries with a given point group. A number of Higgs doublets greater than one are counted as exotics in the average numbers. For scalars, $(\rep3,\rep1)_{\nicefrac{1}{6}} \equiv (\crep3,\rep1)_{\nicefrac{-1}{6}}$.}   
\label{tab:AN-es}
\end{center}
\end{table}
\end{landscape}

\begin{table}[p]
\begin{center}
\resizebox{\textwidth}{!}{
\begin{tabular}{|c|r|r|r|r|r|r|r|r|r|r|r|r|r|r|r|}
\hline
                      & \Z3 & \Z4 & \Z6-I & \Z6-II & \Z8-I & \Z8-II & \Z{12}-II & \Z2\x\Z2 & \Z2\x\Z4 & \Z2\x\Z6-I & \Z2\x\Z6-II & \Z3\x\Z3 & \Z3\x\Z6 
                      & \Z4\x\Z4 & \Z6\x\Z6 \\ \hline
\# SM   & 155 & 30 & 64 & 260 & 8 & 423 & 102 & 234 & 147,996 & 606 & 457 & 14,891 & 147 & 4,834 & 12   \\ \hline\hline
1  & 0\%     & 0\%     & 20.31\% & 0\%     & 0\%     & 0\%     & 0\%     & 0\%     & 0\%     & 0\%     & 0\%     & 0\%     & 0\%     & 0\%     & 0\%     \\ 
2  & 0\%     & 0\%     & 0\%     & 0\%     & 0\%     & 1.42\%  & 0\%     & 0\%     & 2.14\%  & 0.33\%  & 0\%     & 0.07\%  & 0\%     & 0\%     & 0\%     \\ 
3  & 7.10\%  & 0\%     & 0\%     & 6.54\%  & 0\%     & 1.89\%  & 2.94\%  & 0\%     & 0.33\%  & 4.46\%  & 0\%     & 21.04\% & 0\%     & 0.14\%  & 0\%     \\ 
4  & 0\%     & 6.67\%  & 45.31\% & 5.00\%  & 0\%     & 1.89\%  & 7.84\%  & {0.43}\%  & 1.75\%  & 1.82\%  & 0\%     & 0\%     & 0\%     & 0.31\%  & 0\%     \\ 
5  & 0\%     & 0\%     & 0\%     & 5.00\%  & 0\%     & 8.27\%  & 21.57\% & 0\%     & 2.16\%  & 3.30\%  & 1.97\%  & 0.77\%  & 0\%     & 1.37\%  & 0\%     \\ 
6  & 30.32\% & 46.67\% & 25.00\% & 6.15\%  & 0\%     & 7.57\%  & 27.45\% & {8.55}\%  & 4.82\%  & 2.48\%  & 0\%     & 43.44\% & {2.72}\%  & 1.88\%  & 0\%     \\ 
7  & 0\%     & 0\%     & 0\%     & 4.62\%  & 0\%     & 17.26\% & 8.82\%  & {1.28}\%  & 5.58\%  & 5.28\%  & 11.60\% & 0.40\%  & 0.68\%  & 3.19\%  & 8.33\%  \\ 
8  & 0\%     & 46.67\% & 0\%     & 6.92\%  & 0\%     & 9.93\%  & 6.86\%  & {8.55}\%  & 5.08\%  & 4.62\%  & 24.51\% & 1.71\%  & 0\%     & 4.92\%  & 0\%     \\ 
9  & 48.39\% & 0\%     & 3.13\%  & 13.85\% & 0\%     & 17.73\% & 9.80\%  & {0.85}\%  & 7.64\%  & 10.40\% & 10.07\% & 31.40\% & {6.80}\%  & 7.41\%  & 0\%     \\ 
10 & 0\%     & 0\%     & 3.13\%  & 4.23\%  & 0\%     & 6.15\%  & 5.88\%  & {13.25}\% & 5.55\%  & 5.12\%  & 10.72\% & 0.06\%  & {20.41}\% & 9.27\%  & 0\%     \\ 
11 & 1.29\%  & 0\%     & 3.13\%  & 4.23\%  & 0\%     & 9.69\%  & 3.92\%  & {7.26}\%  & 13.31\% & 5.28\%  & 3.06\%  & 0.39\%  & {6.80}\%  & 10.07\% & 16.67\% \\ 
12 & 0.65\%  & 0\%     & 0\%     & 6.54\%  & 0\%     & 3.31\%  & 0.98\%  & {16.67}\% & 6.48\%  & 3.47\%  & 12.69\% & 0.28\%  & {2.72}\%  & 9.97\%  & 8.33\%  \\ 
13 & 0\%     & 0\%     & 0\%     & 9.23\%  & 25.00\% & 5.67\%  & 0\%     & {11.54}\% & 9.08\%  & 8.42\%  & 13.79\% & 0.01\%  & {5.44}\%  & 9.58\%  & 0\%     \\ 
14 & 0\%     & 0\%     & 0\%     & 5.77\%  & 0\%     & 0.47\%  & 1.96\%  & {11.11}\% & 3.38\%  & 2.64\%  & 4.38\%  & 0\%     & {1.36}\%  & 8.46\%  & 8.33\%  \\ 
15 & 9.68\%  & 0\%     & 0\%     & 6.15\%  & 50.00\% & 6.86\%  & 1.96\%  & {12.39}\% & 10.77\% & 2.81\%  & 3.06\%  & 0.21\%  & {8.16}\%  & 7.78\%  & 8.33\%  \\ 
16 & 0\%     & 0\%     & 0\%     & 2.69\%  & 0\%     & 0.47\%  & 0\%     & {2.14}\%  & 4.26\%  & 2.31\%  & 1.31\%  & 0\%     & {2.04}\%  & 7.12\%  & 25.00\% \\ 
17 & 0\%     & 0\%     & 0\%     & 3.08\%  & 0\%     & 0.71\%  & 0\%     & {4.70}\%  & 3.55\%  & 2.64\%  & 0.66\%  & 0.06\%  & {1.36}\%  & 4.70\%  & 0\%     \\ 
18 & 0\%     & 0\%     & 0\%     & 4.23\%  & 0\%     & 0.24\%  & 0\%     & {0}\%  & 1.59\%  & 16.34\% & 0.44\%  & 0\%     & 0\%     & 3.56\%  & 0\%     \\ 
19 & 0\%     & 0\%     & 0\%     & 0.38\%  & 25.00\% & 0\%     & 0\%     & {0.85}\%  & 5.42\%  & 0.66\%  & 1.53\%  & 0\%     & {4.08}\%  & 3.81\%  & 0\%     \\ 
20 & 0\%     & 0\%     & 0\%     & 0\%     & 0\%     & 0\%     & 0\%     & {0}\%  & 1.84\%  & 0.99\%  & 0\%     & 0\%     & 0.68\%  & 2.11\%  & 0\%     \\ 
21 & 1.94\%  & 0\%     & 0\%     & 1.54\%  & 0\%     & 0\%     & 0\%     & {0.43}\%  & 1.36\%  & 1.16\%  & 0\%     & 0.06\%  & {5.44}\%  & 1.97\%  & 8.33\%  \\ 
22 & 0\%     & 0\%     & 0\%     & 0.38\%  & 0\%     & 0\%     & 0\%     & 0\%     & 0.19\%  & 9.41\%  & 0\%     & 0\%     & 0\%     & 0.70\%  & 0\%     \\ 
23 & 0\%     & 0\%     & 0\%     & 0\%     & 0\%     & 0.47\%  & 0\%     & 0\%     & 2.28\%  & 0.83\%  & 0\%     & 0.05\%  & 0\%     & 0.56\%  & 16.67\% \\ 
24 & 0\%     & 0\%     & 0\%     & 1.54\%  & 0\%     & 0\%     & 0\%     & 0\%     & 0.39\%  & 0\%     & 0\%     & 0\%     & {2.04}\%  & 0.29\%  & 0\%     \\ 
25 & 0\%     & 0\%     & 0\%     & 0.38\%  & 0\%     & 0\%     & 0\%     & 0\%     & 0.13\%  & 0.17\%  & 0.22\%  & 0\%     & {5.44}\%  & 0.29\%  & 0\%     \\ 
26 & 0\%     & 0\%     & 0\%     & 0\%     & 0\%     & 0\%     & 0\%     & 0\%     & 0.05\%  & 0\%     & 0\%     & 0\%     & {3.40}\%  & 0.21\%  & 0\%     \\ 
27 & 0.65\%  & 0\%     & 0\%     & 1.15\%  & 0\%     & 0\%     & 0\%     & 0\%     & 0.78\%  & 0.17\%  & 0\%     & 0.03\%  & {4.76}\%  & 0.21\%  & 0\%     \\ 
28 & 0\%     & 0\%     & 0\%     & 0\%     & 0\%     & 0\%     & 0\%     & {0}\%  & 0.03\%  & 1.82\%  & 0\%     & 0\%     & 0\%     & 0.08\%  & 0\%     \\ 
29 & 0\%     & 0\%     & 0\%     & 0\%     & 0\%     & 0\%     & 0\%     & 0\%     & 0.01\%  & 0.50\%  & 0\%     & 0.01\%  & {2.04}\%  & 0.02\%  & 0\%     \\ 
30 & 0\%     & 0\%     & 0\%     & 0.38\%  & 0\%     & 0\%     & 0\%     & 0\%     & 0.01\%  & 0\%     & 0\%     & 0\%     & 0.68\%  & 0\%     & 0\%     \\ 
31 & 0\%     & 0\%     & 0\%     & 0\%     & 0\%     & 0\%     & 0\%     & 0\%     & 0.03\%  & 0.17\%  & 0\%     & 0\%     & {2.04}\%  & 0\%     & 0\%     \\ 
32 & 0\%     & 0\%     & 0\%     & 0\%     & 0\%     & 0\%     & 0\%     & 0\%     & 0.01\%  & 0.50\%  & 0\%     & 0\%     & 0\%     & 0\%     & 0\%     \\ 
33 & 0\%     & 0\%     & 0\%     & 0\%     & 0\%     & 0\%     & 0\%     & 0\%     & 0.01\%  & 0\%     & 0\%     & 0.01\%  & {2.04}\%  & 0.02\%  & 0\%     \\ 
35 & 0\%     & 0\%     & 0\%     & 0\%     & 0\%     & 0\%     & 0\%     & 0\%     & 0.01\%  & 0.17\%  & 0\%     & 0.01\%  & {1.36}\%  & 0\%     & 0\%     \\ 
36 & 0\%     & 0\%     & 0\%     & 0\%     & 0\%     & 0\%     & 0\%     & 0\%     & 0\%     & 0.33\%  & 0\%     & 0\%     & 0\%     & 0\%     & 0\%     \\ 
37 & 0\%     & 0\%     & 0\%     & 0\%     & 0\%     & 0\%     & 0\%     & 0\%     & 0.01\%  & 0\%     & 0\%     & 0\%     & 0.68\%  & 0\%     & 0\%     \\ 
38 & 0\%     & 0\%     & 0\%     & 0\%     & 0\%     & 0\%     & 0\%     & 0\%     & 0.01\%  & 0\%     & 0\%     & 0\%     & {1.36}\%  & 0\%     & 0\%     \\ 
39 & 0\%     & 0\%     & 0\%     & 0\%     & 0\%     & 0\%     & 0\%     & 0\%     & 0.01\%  & 0.17\%  & 0\%     & 0\%     & 0.68\%  & 0\%     & 0\%     \\ 
41 & 0\%     & 0\%     & 0\%     & 0\%     & 0\%     & 0\%     & 0\%     & 0\%     & 0\%     & 0\%     & 0\%     & 0\%     & {2.72}\%  & 0\%     & 0\%     \\ 
43 & 0\%     & 0\%     & 0\%     & 0\%     & 0\%     & 0\%     & 0\%     & 0\%     & 0\%     & 0\%     & 0\%     & 0\%     & {1.36}\%  & 0\%     & 0\%     \\ 
44 & 0\%     & 0\%     & 0\%     & 0\%     & 0\%     & 0\%     & 0\%     & 0\%     & 0\%     & 0.99\%  & 0\%     & 0\%     & 0\%     & 0\%     & 0\%     \\ 
48 & 0\%     & 0\%     & 0\%     & 0\%     & 0\%     & 0\%     & 0\%     & 0\%     & 0\%     & 0.17\%  & 0\%     & 0\%     & 0\%     & 0\%     & 0\%     \\ 
49 & 0\%     & 0\%     & 0\%     & 0\%     & 0\%     & 0\%     & 0\%     & 0\%     & 0\%     & 0\%     & 0\%     & 0\%     & 0.68\%  & 0\%     & 0\%     \\ 
55 & 0\%     & 0\%     & 0\%     & 0\%     & 0\%     & 0\%     & 0\%     & 0\%     & 0\%     & 0.17\%  & 0\%     & 0\%     & 0\%     & 0.02\%  & 0\%     \\ \hline
\end{tabular}
} 
\caption{Percentage of SM-like models with a certain number of Higgs doublets. We label in the 
first row the orbifold point groups considered, such that the second row provides the total number 
of SM-like models found with all orbifold geometries associated with the indicated point groups. In 
the first column we present the possible number of Higgs doublets in our SM-like models. Models 
with one Higgs doublet were found only in the \Z6-I orbifold.}   
\label{tab:numbH}
\end{center}
\end{table}

\clearpage
\begin{landscape}
\section{Features of \emph{almost SM} models} 
\label{app:ASMtables}

\begin{table}[h!]
\begin{center}
\resizebox{1.5\textwidth}{!}{
\begin{tabular}{|ll|r|r|r|r|r|r|r|r|r|r|r|r|r|r|r|}
\cline{3-15}
\multicolumn{2}{r}{} & \multicolumn{13}{|c|}{{\it almost SM}} & \multicolumn{2}{r}{} \\
\hline
\multicolumn{2}{|r|}{}   & \multicolumn{3}{c|}{models without exotic fermions}    & \multicolumn{10}{c|}{models without exotic scalars} & \multicolumn{2}{c|}{models with 1 Higgs}  \\
\multicolumn{2}{|r|}{} & $\Z2\x\Z4\,(1,6)$   & $\Z2\x\Z4\,(2,4)$ & $\Z3\x\Z3\,(1,4)$ & $\Z2\x\Z4\,(1,5)$ & $\Z2\x\Z4\,(1,6)$ & $\Z2\x\Z4\,(2,3)$ & $\Z2\x\Z4\,(2,4)$ & $\Z2\x\Z4\,(3,5)$ & $\Z2\x\Z4\,(3,6)$ & $\Z2\x\Z4\,(4,3)$ & $\Z2\x\Z4\,(4,4)$ & $\Z2\x\Z4\,(6,5)$ & $\Z2\x\Z4\,(8,3)$ &
$\Z6\text{-I}\,(1,1)$ & $\Z6\text{-I}\,(2,1)$ \\ \hline
\multicolumn{2}{|c|}{\#SM}            &  13   & 21 & 11 & 3 & 117 & 10 & 129 & 1 & 82 & 1 & 46 & 80 & 33 & 7 & 6 \\\hline
\multicolumn{2}{|c|}{\#Higgses}       & $\geq 8$ & $\geq 6$ & 9 & 6 & 6 & 6 & 6 & 6 & 6 & 6 & 6 & 6 & 6 & 1 & 1 \\\hline\hline
\multirow{9}{*}{\rotatebox[origin=c]{90}{VLE fermions}} & $(\rep3,\rep2)_{\nicefrac{1}{6}}$             & 0      & 0      & 0      & 0      & 0      & 0      & 0      & 0      & 0      & 0      & 0      & 0      & 0      & 0      & 0      \\
& $(\overline{\rep3},\rep1)_{\nicefrac{-2}{3}}$ & 0      & 0      & 0      & 0      & 0      & 0      & 0      & 0      & 0      & 0      & 0      & 0      & 0      & 0      & 0      \\
& $(\overline{\rep3},\rep1)_{\nicefrac{1}{3}}$  & 0      & 0      & 0      & 6.33   & 6.96   & 8.20   & 7.60   & 4.00   & 5.46   & 7.00   & 8.17   & 5.43   & 5.27   & 5.29   & 5.33   \\
& $(\rep1,\rep2)_{\nicefrac{-1}{2}}$            & 0      & 0      & 0      & 7.67   & 6.74   & 7.40   & 7.49   & 6.00   & 5.71   & 7.00   & 7.91   & 5.68   & 5.70   & 3.57   & 3.50   \\
& $(\rep1,\rep1)_{1}$               & 0      & 0      & 0      & 0      & 0      & 0      & 0      & 0      & 0      & 0      & 0      & 0      & 0      & 0      & 0      \\
& $(\rep1,\rep1)_{0}$               & 130.46 & 111.43 & 120.36 & 168.67 & 176.79 & 188.40 & 186.64 & 200.00 & 159.12 & 210.00 & 208.00 & 159.15 & 166.91 & 197.43 & 201.33 \\
& $(\rep3,\rep1)_{\nicefrac{1}{6}}$             & 0      & 0      & 0      & 0      & 0      & 0      & 0      & 0      & 0      & 0      & 0      & 0      & 0      & 4.71   & 2.17   \\
& $(\rep1,\rep2)_{0}$               & 0      & 0      & 0      & 0      & 0      & 0      & 0      & 0      & 0      & 0      & 0      & 0      & 0      & 7.14   & 4.67   \\
& $(\rep1,\rep1)_{\nicefrac{1}{2}}$             & 0      & 0      & 0      & 0      & 0      & 0      & 0      & 0      & 0      & 0      & 0      & 0      & 0      & 28.14  & 16.83  \\ \hline
\multirow{6}{*}{\rotatebox[origin=c]{90}{Exotic scalars}} & $(\rep1,\rep2)_{\nicefrac{1}{2}}$             & 12.85  & 10.33  & 8.00   & 5.00   & 5.00   & 5.00   & 5.00   & 5.00   & 5.00   & 5.00   & 5.00   & 5.00   & 5.00   & 0      & 0      \\
& $(\rep1,\rep1)_{0}$               & 137.23 & 111.43 & 30.00  & 117.33 & 130.31 & 129.60 & 132.56 & 136.00 & 117.46 & 152.00 & 153.22 & 117.25 & 124.61 & 71.29  & 71.00  \\
& $(\overline{\rep3},\rep1)_{\nicefrac{1}{3}}$  & 11.54  & 10.29  & 9.00   & 0      & 0      & 0      & 0      & 0      & 0      & 0      & 0      & 0      & 0      & 2.00   & 2.00   \\
& $(\rep3,\rep1)_{\nicefrac{1}{6}}$             & 0      & 0      & 0      & 0      & 0      & 0      & 0      & 0      & 0      & 0      & 0      & 0      & 0      & 9.14   & 7.00   \\
& $(\rep1,\rep2)_{0}$               & 0      & 0      & 0      & 0      & 0      & 0      & 0      & 0      & 0      & 0      & 0      & 0      & 0      & 17.29  & 12.00  \\
& $(\rep1,\rep1)_{\nicefrac{1}{2}}$             & 0      & 0      & 0      & 0      & 0      & 0      & 0      & 0      & 0      & 0      & 0      & 0      & 0      & 66.57  & 39.00 \\ \hline
\end{tabular}
}
\caption{Average numbers of vector-like exotic (VLE) fermions and scalars for the different 
categories of {\it almost SM} models, defined in section~\ref{sec:almostSM}. We show in the 
header the categories and the orbifold geometries where the models are found. \#SM indicates the 
number of models identified. \#Higgses indicates the number of Higgs doublets, including the SM 
Higgs. To compute the averages, we consider that all extra Higgses are exotic scalars. For 
comparison, we provide in the last two columns the properties of all 13 SM-like models 
exhibiting just one Higgs doublet.}
\label{tab:AN-ASM}
\end{center}
\end{table}

\enlargethispage{\baselineskip}

\begin{table}[h!]
\begin{center}
\begin{tabular}{|c||r|r|r|r|r|}
\hline
Orbifold       & \multicolumn{5}{c|}{\# of {\it almost SM} models with $n$ Higgs doublets} \\
geometry       & $n ~=~$ 6  & \phantom{$n ~=~$} 8  & \phantom{$n ~=~$} 10  & \phantom{$n ~=~$} 14  & \phantom{$n ~=~$} 18 \\
\hline\hline
\Z2\x\Z4 (1,6) &  0  &  1  &  3   &  5   &  4  \\
\Z2\x\Z4 (2,4) &  3  &  0  &  9   &  8   &  1  \\
\hline
\end{tabular}
\caption{Number of {\it almost SM} orbifold models without exotic fermions (see table~\ref{tab:AN-ASM})
with various numbers of Higgs doublets, including the SM Higgs.}
\label{tab:numbH-ASM}
\end{center}
\end{table}

\end{landscape}

\newpage

\section{Correlations in the landscape of SM-like string models}
\label{app:correlations}

\begin{figure}[h!]
\begin{center}
\begin{tabular}{cc}
\includegraphics[width=0.5\linewidth]{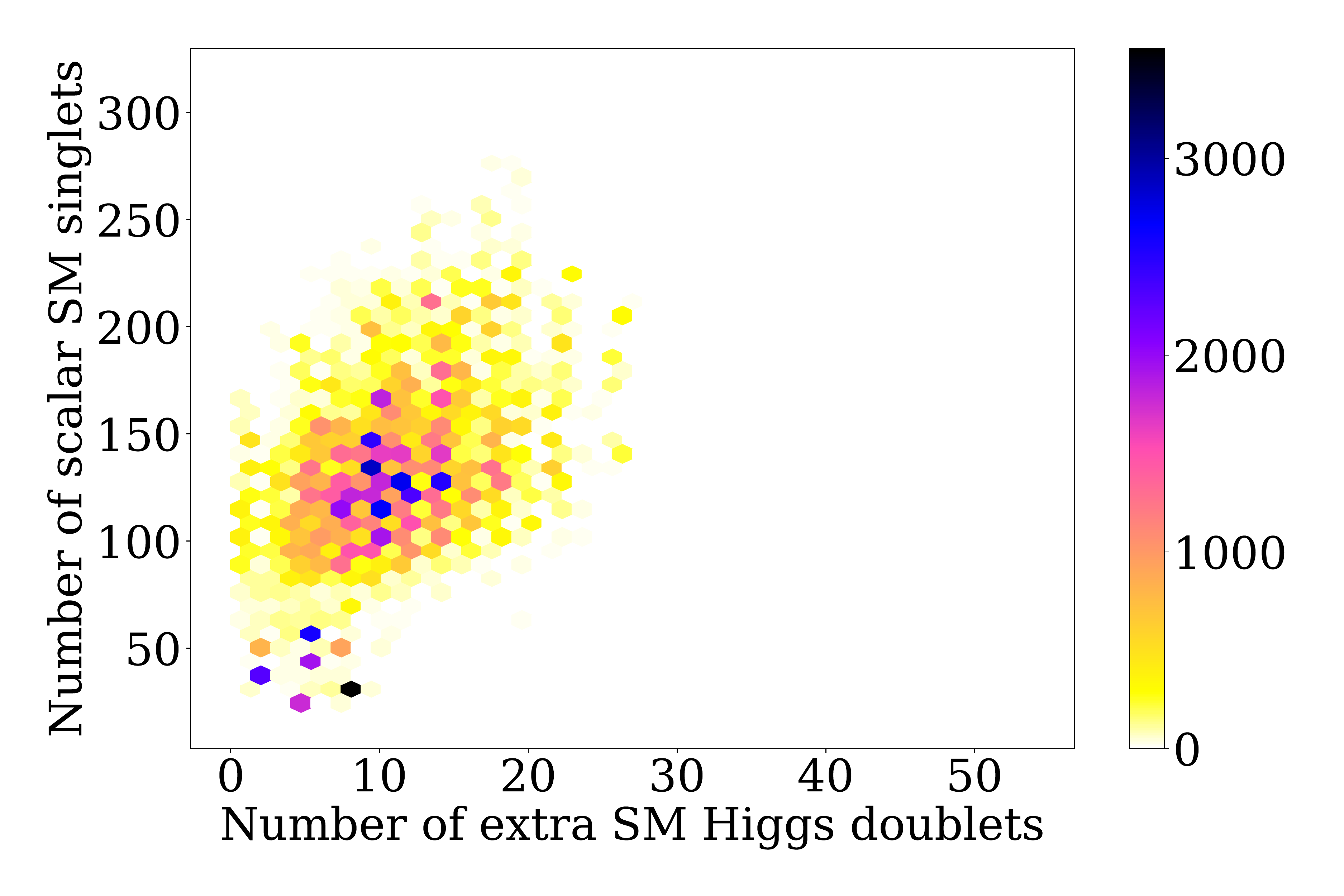}             & \includegraphics[width=0.5\linewidth]{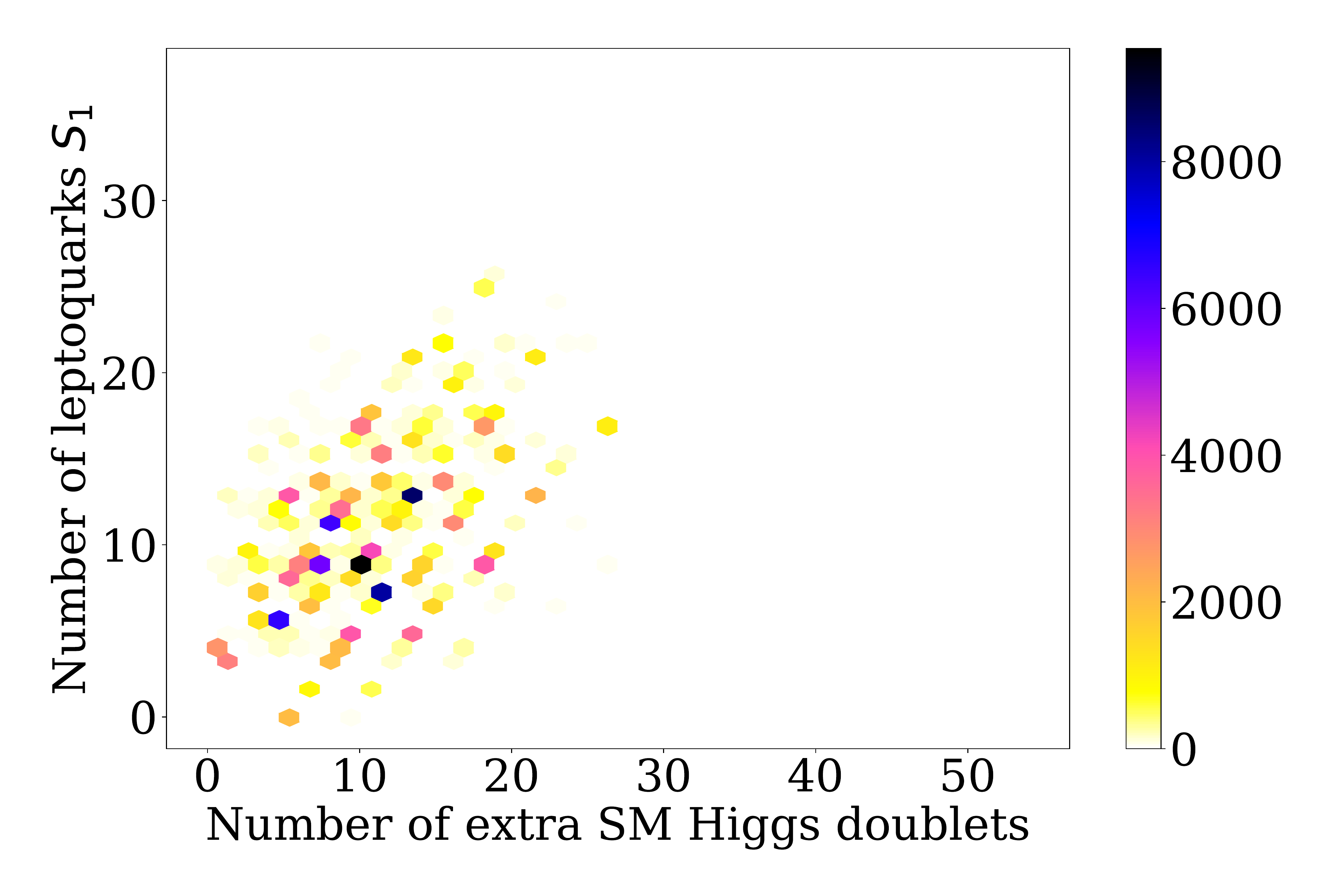}\\
\includegraphics[width=0.5\linewidth]{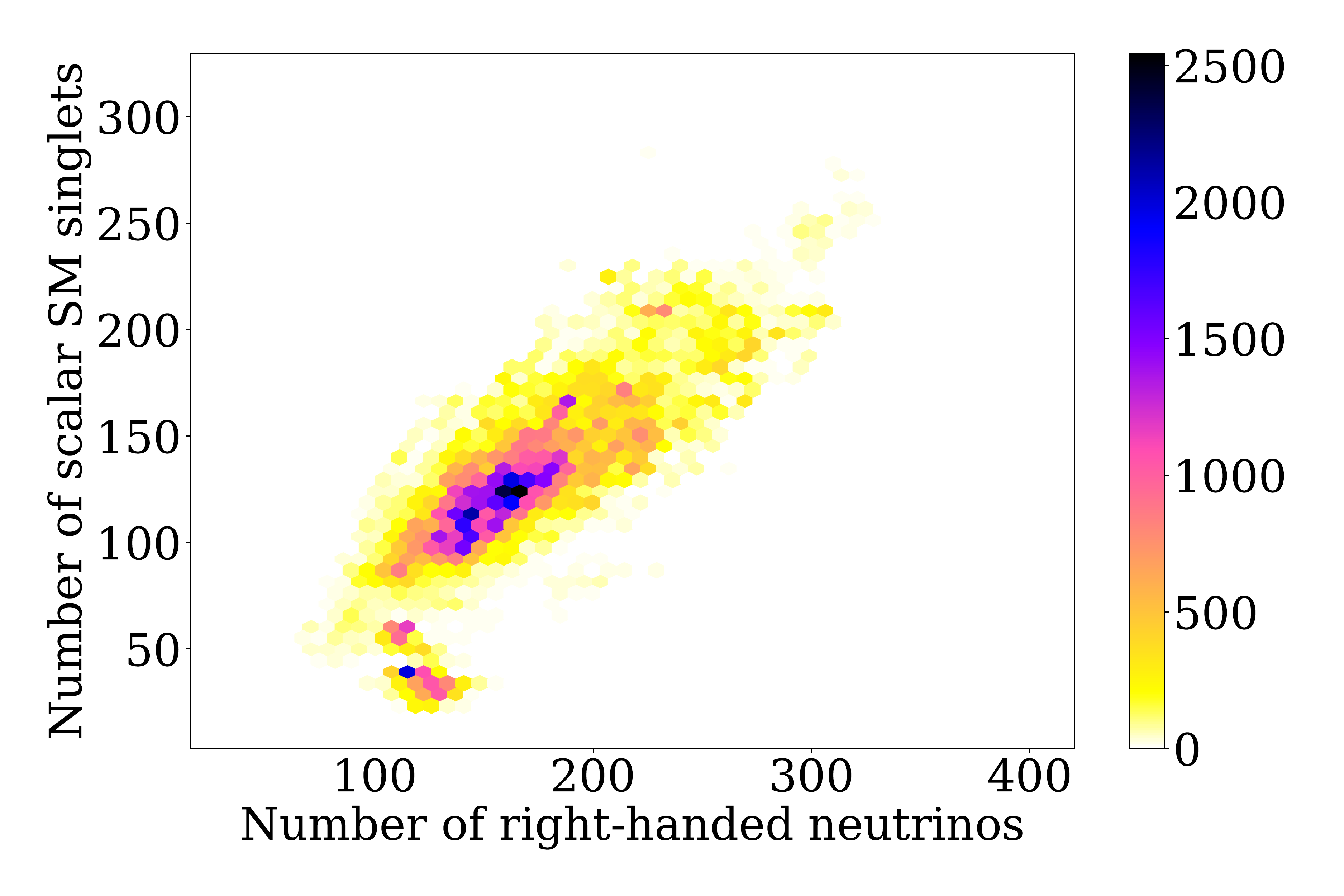}       & \includegraphics[width=0.5\linewidth]{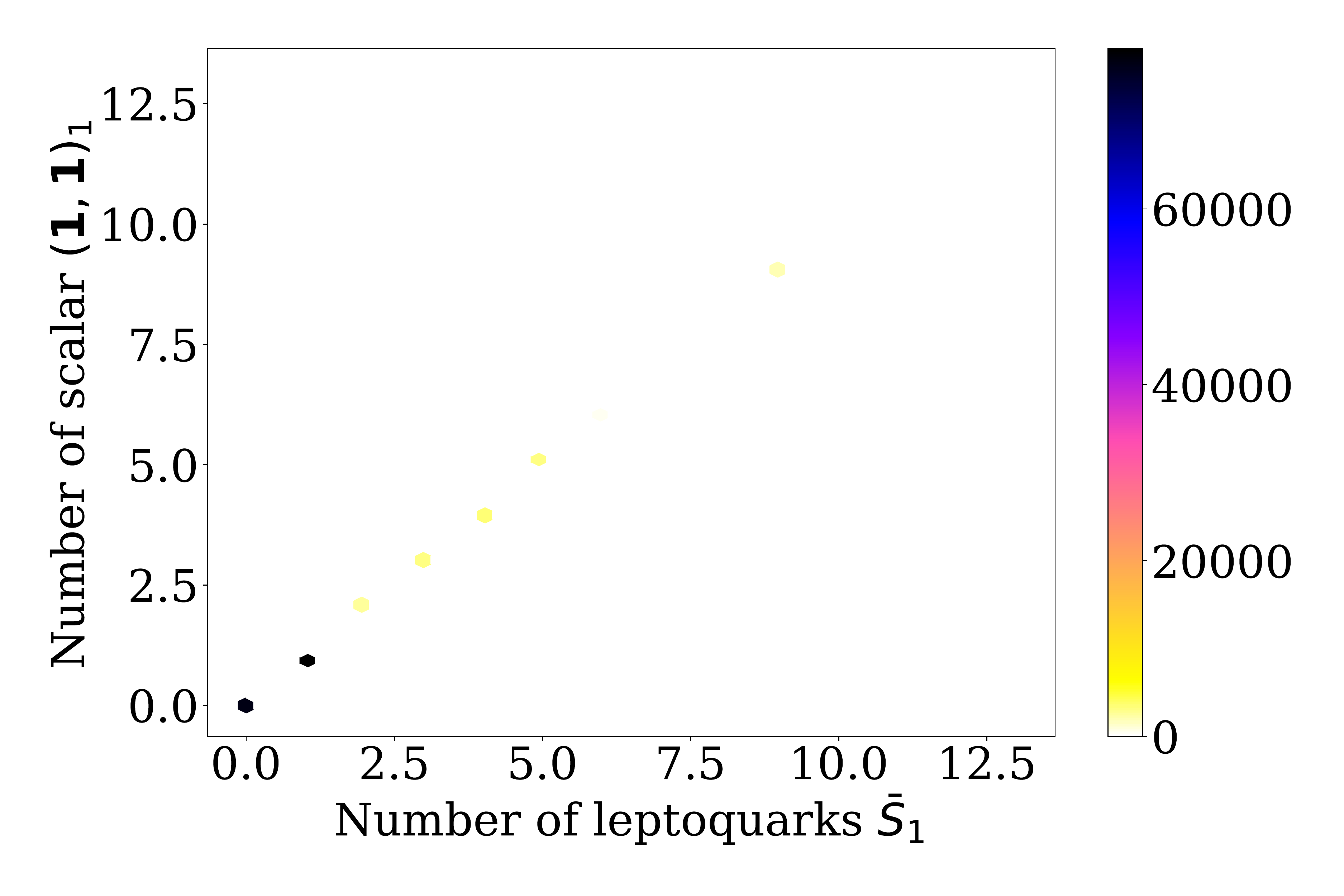}\\
\includegraphics[width=0.5\linewidth]{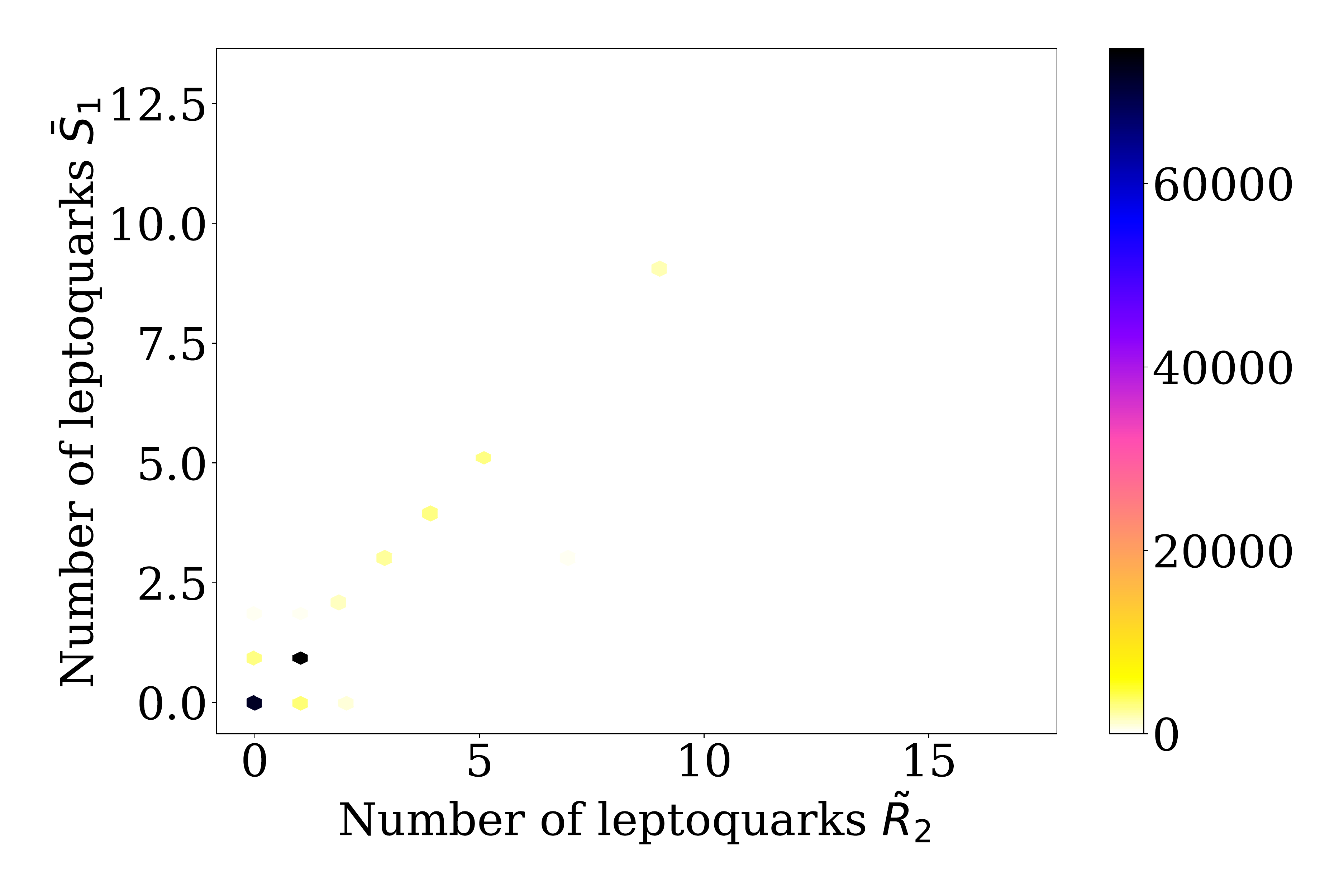} & \includegraphics[width=0.5\linewidth]{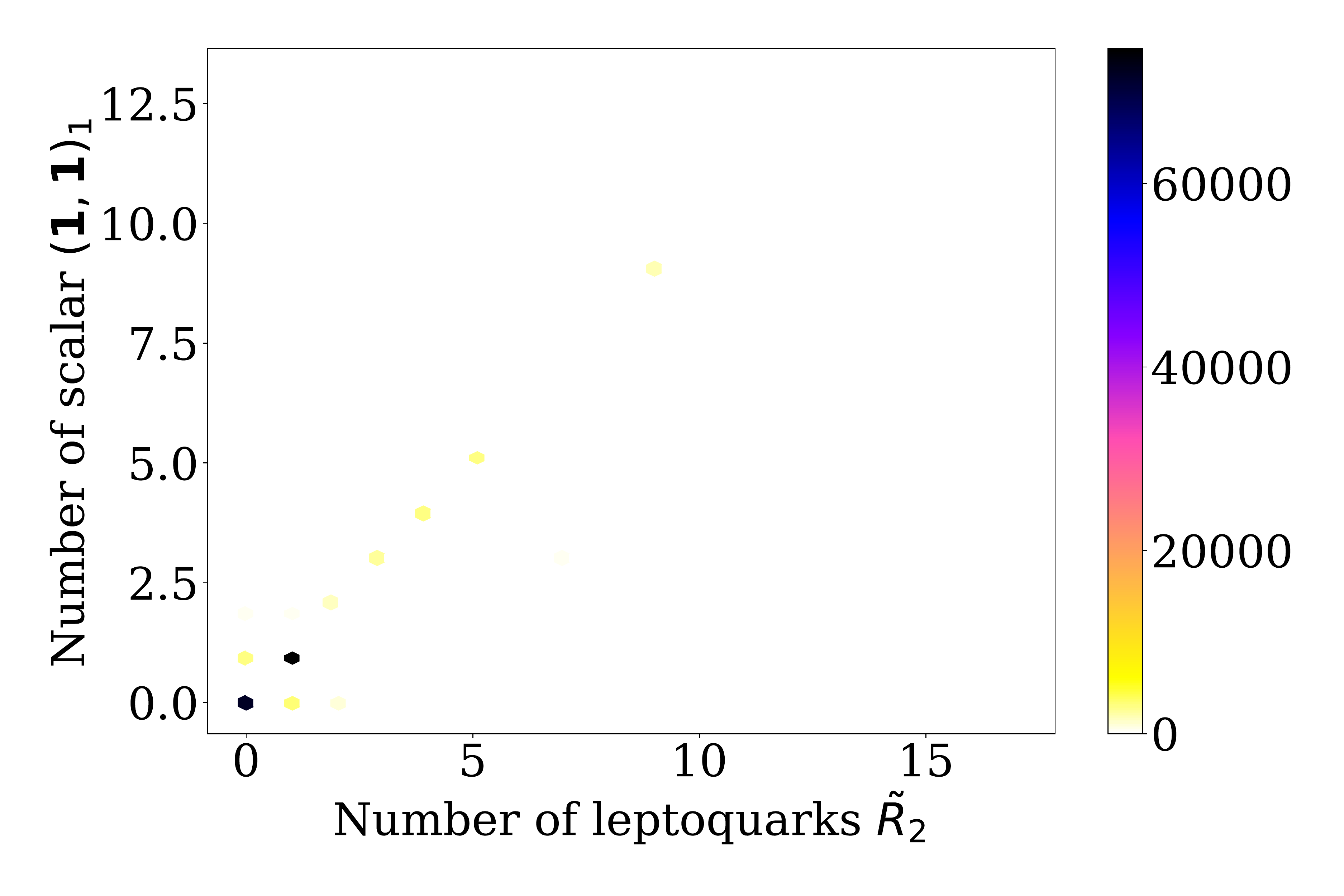}\\
\end{tabular}
\end{center}
\vspace{-0.8cm}
\caption{Some (high) correlations between numbers of exotics in our dataset of 170,219 SM-like 
string models. The correlations are computed on the full dataset and they read from left to right 
and top to bottom: $0.49$, $0.49$, $0.81$, $0.99$, $0.91$ and $0.91$. They point towards the 
existence of local GUTs~\cite{Forste:2004ie,Buchmuller:2004hv,Ratz:2007my,Nilles:2014owa}, where at 
certain orbifold singularities complete GUT representations are localized even though the 
four-dimensional gauge group is just $\mathcal{G}_{\mathrm{SM}}$. The correlation between the 
number of extra Higgs doublets and the number of scalar color triplets $S_1$ seems to originate 
from complete scalar $\rep{5}$-plets of local $\SU{5}$ GUTs, while the correlations between the 
number of Higgs doublets and the scalar leptoquarks $\bar{S}_1$ and $\tilde{R}_2$ might result from 
complete scalar $\rep{10}$-plets of \SU5.}
\label{fig:correlations}
\end{figure}

\end{appendix}

{\small
\providecommand{\bysame}{\leavevmode\hbox to3em{\hrulefill}\thinspace}

}

\end{document}